\documentclass{article}
\usepackage{arxiv}   
\usepackage{natbib}  
\usepackage{amsmath,amssymb}
\usepackage{booktabs}
\usepackage{multirow}
\usepackage{microtype}
\usepackage{hyperref}
\usepackage{url}
\usepackage[table]{xcolor}
\usepackage{tikz}
\usetikzlibrary{positioning,arrows.meta,fit,backgrounds}
\usepackage{pgfplots}
\pgfplotsset{compat=1.17}
\usepackage{listings}
\usepackage{enumitem}
\usepackage{etoolbox}
\setlist{nosep,leftmargin=1.4em}
\definecolor{cspec}{HTML}{2F6DAE}
\definecolor{cfree}{HTML}{B84A46}
\definecolor{cplain}{HTML}{6B7280}
\definecolor{cfreeplus}{HTML}{D9822B}
\definecolor{codebg}{HTML}{F8FAFC}
\definecolor{codeframe}{HTML}{D5DEE8}
\definecolor{codekw}{HTML}{1D4F7A}
\definecolor{codecmt}{HTML}{667085}
\definecolor{codestr}{HTML}{9A5B13}
\definecolor{tablerow}{HTML}{F3F6FA}
\definecolor{tableline}{HTML}{606B77}
\definecolor{plotgrid}{HTML}{E6EAF0}
\definecolor{plotaxis}{HTML}{4B5563}
\definecolor{clink}{HTML}{1565C0}
\definecolor{ccite}{HTML}{1E8E3E}
\definecolor{curl}{HTML}{9C27B0}
\hypersetup{
  colorlinks=true,
  linkcolor=clink,
  citecolor=ccite,
  urlcolor=curl,
}
\pgfplotsset{
  every axis/.append style={
    axis line style={draw=plotaxis, line width=0.35pt},
    tick style={draw=plotaxis, line width=0.3pt},
    tick label style={font=\footnotesize},
    label style={font=\small},
    major grid style={draw=plotgrid, line width=0.35pt},
    legend style={draw=codeframe, fill=white, fill opacity=0.92, draw opacity=1,
      rounded corners=1pt, inner xsep=4pt, inner ysep=2pt},
  },
  every axis plot/.append style={line width=1.05pt, mark size=1.7pt,
    mark options={solid}},
}
\newbool{tlzebra}
\booltrue{tlzebra}
\makeatletter
\AtBeginEnvironment{tabular}{\ifbool{tlzebra}{\rowcolors{2}{}{tablerow}\arrayrulecolor{tableline}}{}}
\AfterEndEnvironment{tabular}{\arrayrulecolor{black}}
\pretocmd{\@maketitle}{\boolfalse{tlzebra}}{}{}
\apptocmd{\@maketitle}{\booltrue{tlzebra}}{}{}
\makeatother
\newlength{\tblw}
\newcommand{\spec}{\textsc{spec}}
\newcommand{\freeplus}{\textsc{free+}}
\newcommand{\free}{\textsc{free}}
\newcommand{\pp}{\,pp}

\title{Specification Grounding Drives Test Effectiveness for LLM Code}
\renewcommand{\shorttitle}{Spec Grounding Drives Test Effectiveness for LLM Code}
\author{%
  Amin Haeri \\
  Model Development Innovation \\
  TD Bank, Toronto, Canada \\
  \texttt{amin.haeri@td.com}
  \And
  Mahdi Ghelichi \\
  Model Development Innovation \\
  TD Bank, Toronto, Canada \\
  \texttt{mahdi.ghelichi@td.com}
}
\date{}

\begin{document}
\maketitle

\begin{abstract}
Large language models frequently generate code that appears correct on typical inputs
yet fails on edge cases, invalid inputs, and specification-defined corner conditions,
such as a missing input check or an unhandled boundary case. A popular fix
has the model write its own tests and repair until they pass. This helps, but the
source of the gain is unclear. Does it come from the tests merely \emph{existing}, or
from their grounding in a \emph{specification} of what the code should do? We isolate
this factor through a controlled experimental design. Holding fixed everything that
normally varies (test count, which model
writes the tests, and the repair loop), we change a single line of the test-writer's
prompt that controls whether it receives the spec as a checklist of rules. The baseline
is no strawman, since it is already told to probe invalid inputs and edge cases.
Grounding the tests in the spec produces correct code \textbf{+38 percentage points}
more often than this strong baseline on each of three Claude tiers (Haiku~4.5,
Sonnet~4.6, Opus~4.8), and $+36$ points more often on a held-out set. These results
indicate that specification grounding, rather than test quantity, is the primary driver
of test effectiveness. Doubling the baseline's budget barely helps, and even
combining eight independent ungrounded test suites plateaus far below grounding. An
ablation pins the cause to the spec's \emph{content} rather than its format. Given the
spec as a plain paragraph, the tester still recovers 27 of 30 bugs; asked to plan tests
without the spec, it recovers only 2 of 30. The effect remains robust under stronger
baselines. A property-based generator catches 28 of 30 bugs but invents out-of-spec
requirements, and a full AlphaCodium-style loop only matches the baseline. It also holds
across three model families run full-stack. Beyond Claude, GPT-5.3-codex gains $+28$
points and Gemini~3.5~Flash gains $+19$. A task-level sign test over 18 tasks is
significant at $p=0.002$ (only about a 0.2\% chance of a gap this consistent if
grounding made no difference).
Specification grounding improves both sensitivity and precision: more real bugs are
caught \emph{and} far less correct code is wrongly rejected. The false-alarm rate is 0\%
for the grounded tests versus 33\% for
the baseline, rising to 68\% for the baseline when the oracle is the Python standard
library itself. On well-specified algorithmic problems it neither
helps nor hurts.
\end{abstract}

\section{Introduction}

A modern language model, asked to write a small function, almost always returns
something that compiles and handles the common case. Failures frequently arise in
boundary conditions and under-specified behaviours. Consider a one-line ticket: \emph{``parse a compact integer range: `a-b' expands to
the inclusive list \texttt{[a,\dots,b]}; a bare `n' returns \texttt{[n]}.''} Claude
Opus~4.8 writes:

\begin{lstlisting}[language=Python]
def parse_range(spec: str) -> list[int]:
    spec = spec.strip()
    start, sep, end = spec.partition("-")
    if sep == "":              # bare 'n'
        return [int(spec)]
    a, b = int(start), int(end)
    if a > b:
        raise ValueError(f"start {a} > end {b}")
    return list(range(a, b + 1))
\end{lstlisting}

\noindent This is careful code: it strips whitespace, handles the bare \texttt{"n"}
case, and even guards the reversed range, so \texttt{"5-1"} raises a clear error
instead of returning an empty list. Yet corners remain. A malformed \texttt{"1-2-3"},
an empty string, and a negative bound such as \texttt{"-3-5"} all crash with an opaque
\texttt{int()} error rather than a clear ``malformed range'' report; the model itself
flagged the negative case as an unhandled assumption. In our experiments \emph{every}
model we tried, including the strongest, left omissions of exactly this kind. Such
missing input checks are not exotic; they are the everyday gap between ``plausible''
and ``correct,'' and a leading source of real-world defects.

The field's standard response is to generate tests and let the model iterate: run
the tests, feed back the failures, repair. This demonstrably improves code
generation, from test-based selection~\citep{chen2022codet} to flow-based
agents~\citep{ridnik2024alphacodium}. But these methods establish only \emph{that}
AI-written tests help. They leave open \emph{why}, and the answer determines what a
practitioner should actually do.

\paragraph{Two competing explanations.}
Tests might help for either of two reasons, and the two point to opposite advice. The
first is sheer quantity. Running any plausible test sometimes trips a bug, so more tests
catch more, and the advice is simply to generate more of them. The second is grounding.
Tests drawn from an external specification are aimed at the behaviours that matter, above
all the invalid-input and edge cases a model tends to forget, and the advice is to write
a specification and turn each clause into a test. These two hypotheses imply different
practical recommendations: increasing test volume, or investing effort in explicit
specification development. If quantity is the real driver, specifications add little; if
grounding is, they are essential.

\begin{figure}[t]\centering
\begin{tikzpicture}[
  node distance=5mm and 9mm,
  box/.style={draw, rounded corners, align=center, font=\footnotesize, inner sep=3pt, minimum height=8mm, text width=19mm},
  spec/.style={box, draw=cspec, very thick, fill=cspec!6},
  free/.style={box, draw=cfreeplus, very thick, fill=cfreeplus!6},
  >={Stealth[round]}, thick]
\node[box, text width=13mm] (ticket) {Prose\\ticket};
\node[box, right=of ticket] (impl) {Code model:\\one-shot code};
\node[spec, above right=3mm and 11mm of impl] (sgen) {\spec{}: one test\\\emph{per spec rule}};
\node[free, below right=3mm and 11mm of impl] (fgen) {\freeplus{}: $K$ tests,\\``cover the edges''};
\node[box, right=28mm of impl, text width=15mm] (run) {Run tests;\\repair on fail};
\node[box, right=of run, text width=13mm] (final) {Final\\code};
\node[box, draw=black, dashed, fill=black!4, below=6mm of final, text width=20mm] (gold) {Gold oracle\\(independent,\\hidden)};
\draw[->] (ticket)--(impl);
\draw[->] (impl.east) to[out=20,in=180] (sgen.west);
\draw[->] (impl.east) to[out=-20,in=180] (fgen.west);
\draw[->] (sgen.east) to[out=0,in=120] (run.north);
\draw[->] (fgen.east) to[out=0,in=-120] (run.south);
\draw[->] (run)--(final);
\draw[->] (final)--(gold);
\end{tikzpicture}
\caption{\textbf{The experiment in one picture.} A code model writes one implementation
from a prose ticket. A \emph{fixed} tester then writes tests under one of two prompts
differing in a single line: \spec{} gets the spec as a checklist and writes one test per
rule (blue); \freeplus{}, our fair baseline, gets only the ticket but is told to ``cover
the edges'' (orange). The tests drive an identical repair loop, and an independent gold
oracle the models never see judges the final code. Only the test prompt changes across
arms; \S\ref{sec:ablation} pins the effect on its \emph{content}, not its enumeration.}
\label{fig:pipeline}
\end{figure}
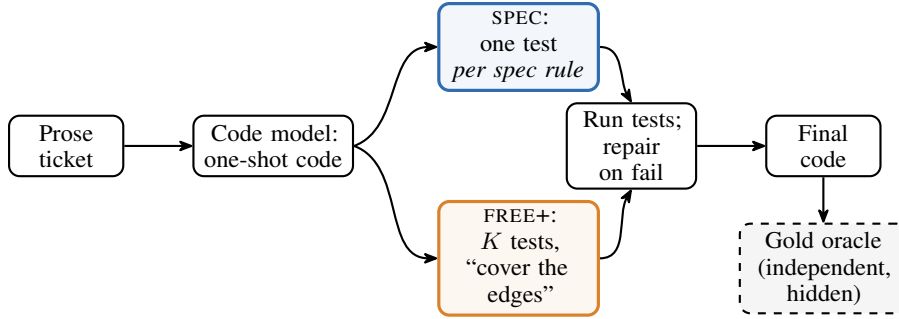

\paragraph{Our experiment.}
We isolate the variable. For each task a code model writes one implementation (the
realistic ``fast pass''; we plant no bugs). A \emph{fixed} tester then writes tests
under one of two prompts that differ in a single line. \spec{} receives the
specification enumerated into discrete rules and writes one test per rule.
\freeplus{} receives only the ticket but is \emph{explicitly told} to test invalid
inputs and edge cases. Tester capability, test budget, and repair procedure are held
constant across conditions. Correctness is
judged not by either set of tests but by an independent, hand-written gold suite
that no model ever sees and that is deliberately broader than the rule list.
Including \freeplus{} is what makes the comparison fair, since it already knows it
should test edges. Any remaining advantage of \spec{} is then about \emph{grounding},
not about reminding a naive tester to think about errors. Figure~\ref{fig:pipeline} shows
the setup.

\noindent We begin with the analysis that most directly identifies the causal mechanism,
explaining \emph{why} grounding helps rather than merely that it does. An ablation
(\S\ref{sec:ablation}) takes the specification apart. A tester given only a plan that
breaks the task into cases, with no spec, catches almost nothing (2 of 30 bugs). Given
the spec as a plain paragraph, it catches nearly all of them (27 of 30). Splitting that
same spec into a checklist and writing one test per rule adds only a little more (30 of
30). So the spec's \emph{content} is the cause, and the main gain of $+38$
percentage points (pp) follows from it. Every cheaper explanation then fails its own
test. Running more ungrounded tests does not close the gap, whether by doubling the
budget or by combining several independent test suites (\S\ref{sec:results},
App.~\ref{app:rw}). Neither does a generic nudge to ``cover the edges'' (the \freeplus{}
baseline), nor a stronger ungrounded generator such as a property-based battery
(\S\ref{sec:prop}) or an AlphaCodium-style flow.

\paragraph{Contributions.}
We identify specification grounding as the principal mechanism through which AI-generated
tests improve code quality, and turn this finding into practical guidance, in five
contributions.
\begin{itemize}[leftmargin=2.2em, itemsep=3pt, topsep=3pt]
\item We isolate grounding from test quantity in LLM test-driven repair, holding the
tester, budget, and repair loop fixed and changing only whether the tester sees the
spec, with correctness judged by an independent gold oracle against a fair,
edge-prompted baseline (\S\ref{sec:method}).
\item We show that grounding helps in two ways at once, catching more real bugs and raising
fewer false alarms, and we trace this to two mechanisms: writing one test per spec
rule surfaces the non-obvious invalid conditions a model forgets, and each rule's failure
gives a complete repair signal (\S\ref{sec:results}).
\item We isolate the contribution of specification \emph{content} from that of task
decomposition and coverage planning, through an ablation that separates content from a
rule-by-rule split and a chain-of-thought plan (\S\ref{sec:ablation}).
\item We bound the claim to specification-completeness defects, not algorithmic logic
(\S\ref{sec:results}; App.~\ref{app:evalplus}).
\item We report a capability-substitution result, where grounding a small model matches
a larger one run alone, with a token-cost accounting (\S\ref{sec:capability};
App.~\ref{app:extra}).
\end{itemize}

\section{Background and related work}\label{sec:related}

The work we build on shares one thesis. Test-driven repair helps LLM code
generation, and what differs across methods is \emph{where the test author's expected
values come from}. We group prior work by the grounding of that signal (ungrounded, self-feedback, and weak
grounding), against the external specification we study; none of these prior groups
isolates grounding as the causal lever, which is our question (Table~\ref{tab:landscape}).

\begin{table}[tb]\centering
\caption{The landscape grouped by \emph{where the test author's expected values come
from}. Only an external specification grounds those expectations, and none of the prior
groups isolates that grounding as the causal lever; \spec{} is our arm. The rows elaborate
the paragraphs of this section.}
\label{tab:landscape}
\small
\setlength{\tblw}{\dimexpr\textwidth-6\tabcolsep\relax}
\begin{tabular}{@{}>{\raggedright\arraybackslash}p{0.40\tblw} >{\raggedright\arraybackslash}p{0.37\tblw} >{\centering\arraybackslash}p{0.13\tblw} >{\centering\arraybackslash}p{0.10\tblw}@{}}
\toprule
Group & Where test expectations come from & Grounding & Isolates grounding? \\
\midrule
Ungrounded test-driven (CodeT, AlphaCodium, LEVER) & Model invents each test's expected value & none & no \\
Self-feedback (Self-Refine, Reflexion, Self-Debugging) & Model critiques its own code & none (self) & no \\
Weak grounding (property-based, TiCoder) & Author invariants, or a human in the loop & weak & no \\
\textbf{External specification (\spec{}, ours)} & \textbf{Enumerated spec rules, one test per rule} & \textbf{external} & \textbf{yes} \\
\bottomrule
\end{tabular}
\end{table}

\paragraph{Ungrounded test-driven generation and repair.}
HumanEval~\citep{chen2021codex} and MBPP~\citep{austin2021mbpp} established
functional-correctness evaluation for code models~\citep{roziere2023codellama};
EvalPlus~\citep{liu2023evalplus} showed thin test suites overstate correctness,
motivating our deliberately broad, independent oracle. On the method side, CodeT
selects code by dual-execution agreement~\citep{chen2022codet}, AlphaCodium couples
generated tests with an iterative flow~\citep{ridnik2024alphacodium}, and LEVER trains
an execution verifier~\citep{ni2023lever}. All treat tests as free-form artifacts whose
expected values the model invents, and none ask whether the benefit is the tests'
\emph{existence or quantity} or their \emph{grounding}. Determining correct expected
outputs remains a manifestation of the ``oracle problem'' in software
testing~\citep{barr2015oracle}. An ungrounded author
must guess, and on non-obvious behaviour it guesses wrong. That is one of the two
mechanisms we measure, and grounding supplies a partial oracle for it.

\paragraph{Self-feedback (self-correction and repair).} Self-Refine~\citep{madaan2023selfrefine},
Reflexion~\citep{shinn2023reflexion}, and Self-Debugging~\citep{chen2024selfdebug}
iterate on feedback the model generates about itself; classical program repair predates
LLMs~\citep{legoues2019apr}. Two results sharpen our motivation. \citet{huang2024selfcorrect}
show intrinsic self-correction often fails or hurts, and \citet{olausson2024selfrepair}
pinpoint poor self-feedback as the bottleneck. An enumerated specification can be viewed
as a structured form of the stronger, \emph{external} feedback advocated in that prior
work, applied at the test-authoring step, and our blind self-refine control confirms it
does not substitute.

\paragraph{Weak grounding (property-based testing and TiCoder).} Writing behaviour down is old wisdom,
from Design by Contract~\citep{meyer1992dbc} and Dafny~\citep{leino2010dafny} to
property-based testing~\citep{claessen2000quickcheck}. The last is a spec-to-test
pipeline we compare against directly (\S\ref{sec:prop}) and find only \emph{weakly}
grounded: a random-plus-invariant generator catches many bugs but \emph{invents}
out-of-spec assertions, the same hallucination an ungrounded edge tester makes. TiCoder formalizes
intent through tests with a human in the loop~\citep{lahiri2022ticoder}; we are fully
automated and isolate the marginal value of \emph{enumerating} the spec for the tester.
We reuse the grounded test-pass fraction as a selective-prediction
signal~\citep{geifman2017selective}, and our independent oracle avoids the
self-enhancement bias of LLM judges~\citep{zheng2023judge}. Finally, agentic systems on
SWE-bench~\citep{jimenez2024swebench} like SWE-agent~\citep{yang2024sweagent} already
iterate against tests; we suggest \emph{where} the tests' expectations come from is a
lever orthogonal to better search. A recent \emph{code-as-harness} survey names
\emph{oracle adequacy}, that verification is only as good as the oracle behind it, as a
key open problem~\citep{ning2026codeharness}; we answer one slice of it.

\paragraph{External specification (our position).} Prior work improves LLM coding through better search, more
tests, or self-feedback; across all three the load-bearing quantity is the
\emph{quality of the feedback signal}. We provide a controlled isolation of its
\emph{grounding in an external specification} as the driver, using a fair, edge-prompted
baseline and an independent oracle, and show the effect on both sides of the
confusion matrix.

\section{Method}\label{sec:method}

\subsection{Overview}
We refer to each test-generation strategy as an experimental \emph{arm}. For each task, a
code model writes one
implementation. An arm then writes tests for that code, the tests drive a repair loop,
and an independent oracle issues the final verdict (Figure~\ref{fig:pipeline}).
Everything is held equal across arms except the test prompt. The two main arms are
\spec{} and \freeplus{}. \spec{} receives the specification split into separate rules
and writes one test per rule. \freeplus{} is a fair baseline that receives only the
ticket, but is told to test invalid inputs and edge cases. The \spec{} prompt differs
from \freeplus{} in three ways: it adds the content of the spec, splits that content
into separate rules, and asks for one test per rule. The ablation in
\S\ref{sec:ablation} separates these three and finds that the added content is what
carries the effect. We also run several further controls and stronger baselines,
including a doubled test budget, combined test suites, a property-based generator, and
an AlphaCodium-style flow, and introduce each one where it is used. The remaining
subsections describe the tasks and their hidden oracle, the arms, the models and repair
loop, and the metrics.

\subsection{Tasks and oracle}
Our benchmark has 26 small Python tasks in two groups. The main group is 18
\emph{specification-completeness} tasks, whose plain-language tickets hide real edge
cases. It has 8 \emph{core} tasks (semver comparison, money splitting, pagination,
median, interval merging, ellipsis truncation, Roman-numeral parsing, and clamping)
used as the main testbed (\S\ref{sec:primary}--\ref{sec:prop}), 4
\emph{held-out} tasks used only for replication (\S\ref{sec:robust}), and 6
\emph{scale-up} tasks added later to enlarge the sample (\S\ref{sec:scaleup}). The second
group is 8 \emph{logic} tasks, where the hard part is the algorithm rather than input
validation; we include them only as a scope control (\S\ref{sec:scope}). All four
subgroups, with their full rule lists, are defined in Appendix~\ref{app:tasks}.

Each task comes with four parts: a realistic prose ticket (the only input the code model
sees), the specification enumerated into $K$ checkable rules (shown only to \spec{}), a
trusted reference implementation, and a hand-written \emph{gold} test suite. The gold
suite is the oracle that judges final correctness. We validate it against worked
examples, never show it to any model, and make it broader than the rule list, so that a
passing solution cannot just be ``the union of the rules.'' We write our own tasks
instead of reusing a benchmark such as HumanEval+~\citep{liu2023evalplus} because
isolating grounding requires each task to provide \emph{both} an enumerated rule spec
\emph{and} an independent oracle; a benchmark's own tests would serve as the spec and
confound the comparison. Standard algorithmic benchmarks also rarely contain this bug
class, as our HumanEval+ port confirms (App.~\ref{app:evalplus}). We therefore need
under-specified tickets with curated edges to elicit the effect; they are not a setup
chosen to favour grounding. A separate set of six tasks (App.~\ref{app:rw}) makes the same
point in a harder way. Their edge cases come from published standards we did not write
(RFC~791, the CSS colour grammar, the ISO ISBN-10 checksum, and similar), so we could not
have picked the edges to suit the method. The gap there is even bigger: \spec{} catches
every bug ($100\%$) while \freeplus{} catches only $32\%$, a $+68$\pp{} gap. These six tasks
are not part of the specification-completeness set or the logic set above. We mean the
benchmark as a controlled isolation study, not a leaderboard.

\subsection{Arms}\label{sec:arms}
A code model writes each function once, and every arm works on that \emph{same} one-shot
code. Our main comparison is between the two test-writing arms. Both write the same
number of tests, $K$ (the number of spec rules for that task), and differ only in their
prompt. \freeplus{}, our fair baseline, sees the ticket and is told to test invalid
inputs and edge cases. \spec{} sees the ticket \emph{and} the spec as $K$ rules, and
writes one test per rule. We add four additional control arms:
\begin{itemize}[leftmargin=2.2em, itemsep=3pt, topsep=3pt]
\item \free{}: $K$ tests with no edge prompt, the AlphaCodium/CodeT-style
baseline~\citep{chen2022codet,ridnik2024alphacodium}.
\item \textsc{free2k}: \free{} with $2K$ tests, to check whether having \emph{more}
tests is what helps.
\item \textsc{self-refine}: the model re-reads and revises its own code, with no tests at
all.
\item \textsc{oneshot}: no checks at all, as a lower bound.
\end{itemize}
In every arm the tester computes the expected outputs from the ticket or the rules, never
from the candidate code, because a tester that sees buggy code tends to treat the bug as
intended behaviour. Example prompts are in Appendix~\ref{app:prompts}.

\subsection{Models, draws, and repair}
Three roles use a model: the \emph{code model} writes the function, the \emph{test author}
writes the tests, and the \emph{repairer} edits the code from the failing tests. We vary the
code model and hold the other two fixed. The code model runs at three tiers of Anthropic's
Claude family, small (Haiku~4.5), medium (Sonnet~4.6), and large (Opus~4.8), so we can see
the effect across model strengths. The test author and the repairer are always the medium
tier (Sonnet~4.6). Fixing the tester this way keeps tester ability constant across arms, so
any gap comes from the test \emph{prompt}, not from a stronger or weaker tester. We leave
every Claude model at its default reasoning effort, the same in \spec{} and \freeplus{}, so
it cannot affect the gap. This main study is Claude-only; the cross-vendor check in
\S\ref{sec:xvendor} repeats the whole pipeline with OpenAI and Google models in \emph{every}
role (code model, test author, and repairer), not just as the code model, and
Appendix~\ref{app:xvendor} lists the generation settings for all of them.

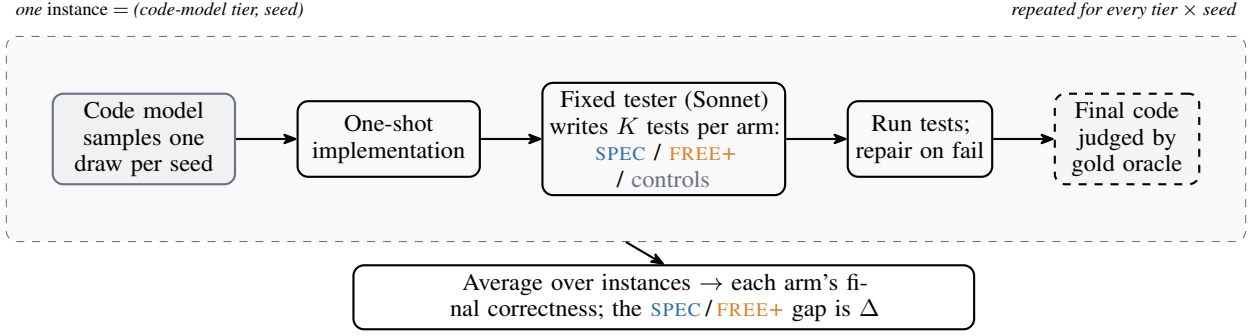
\begin{figure}[t]\centering
\begin{tikzpicture}[
  node distance=6mm and 8mm,
  box/.style={draw, rounded corners, align=center, font=\footnotesize,
              inner sep=3pt, minimum height=10mm, text width=22mm},
  cmbox/.style={box, draw=cplain, fill=cplain!10},
  gold/.style={box, draw=black, dashed, fill=black!4},
  lbl/.style={font=\scriptsize\itshape, align=center},
  >={Stealth[round]}, thick]
\node[cmbox] (cm) {Code model\\samples one\\draw per seed};
\begin{scope}[on background layer]
  \node[cmbox, xshift=3.0mm, yshift=3.0mm] at (cm) {};
  \node[cmbox, xshift=1.5mm, yshift=1.5mm] at (cm) {};
\end{scope}
\node[box, right=of cm] (impl) {One-shot\\implementation};
\node[box, right=of impl, text width=30mm] (tests)
     {Fixed tester (Sonnet)\\writes $K$ tests per arm:\\
      {\color{cspec}\spec} / {\color{cfreeplus}\freeplus} / {\color{cplain}controls}};
\node[box, right=of tests, text width=17mm] (rep) {Run tests;\\repair on fail};
\node[gold, right=of rep, text width=17mm] (final) {Final code\\judged by\\gold oracle};
\draw[->] (cm)--(impl);
\draw[->] (impl)--(tests);
\draw[->] (tests)--(rep);
\draw[->] (rep)--(final);
\begin{scope}[on background layer]
  \node[draw=cplain, dashed, rounded corners, fill=cplain!4, inner sep=6mm,
        fit=(cm)(impl)(tests)(rep)(final)] (plate) {};
\end{scope}
\node[lbl, anchor=south west] at ([yshift=1mm]plate.north west)
     {one \emph{instance} $=$ (code-model tier, seed)};
\node[lbl, anchor=south east] at ([yshift=1mm]plate.north east)
     {repeated for every tier $\times$ seed};
\node[box, below=9mm of tests, text width=80mm, fill=white, minimum height=8mm] (avg)
     {Average over instances $\rightarrow$ each arm's final correctness;
      the {\color{cspec}\spec}\,/\,{\color{cfreeplus}\freeplus} gap is $\Delta$};
\draw[->] (plate.south) -- (avg.north);
\end{tikzpicture}
\caption{\textbf{Draws, seeds, and instances.} The code model is sampled, so each
\emph{seed} is one draw and gives a different one-shot implementation (the grey stack on
the left). One (code-model tier, seed) pair is an \emph{instance} (the dashed plate).
Within an instance every arm tests the \emph{same} implementation and differs only in its
test prompt; the inner arm pipeline is the one in Fig.~\ref{fig:pipeline}. Failing tests
drive an identical repair loop, and a hidden gold oracle judges the final code. Here
\emph{controls} are the other arms of \S\ref{sec:arms}: \free{}, \textsc{free2k}, and the
test-free \textsc{self-refine} and \textsc{oneshot}. We repeat the plate for every tier and
seed and average over instances, so no single lucky or unlucky draw decides a task; the gap
in final correctness between {\color{cspec}\spec} (blue) and {\color{cfreeplus}\freeplus}
(orange) is the main $\Delta$.}
\label{fig:instances}
\end{figure}

The code model samples its output, so the same ticket can give different code each time we
draw from it. We label each draw by a \emph{seed}; a new seed yields a different,
independently generated implementation to put under test. We call one (code-model tier,
seed) pair an \emph{instance} (Figure~\ref{fig:instances}); the tier is the \emph{code model's}, since the test author and
repairer stay at Sonnet throughout. Every arm in an instance is run against that one
implementation, so the arms vary in their test prompts (and, for the quantity control, the
test budget), never in the code under test. The primary
comparison varies the code model over two tiers (Haiku and Sonnet), three seeds
each,\footnote{A \emph{seed} here is a per-draw index, not an API random-seed parameter; the
Anthropic Messages API has none, and we pass no seed to any vendor. We realise the index as
a one-line nonce, \texttt{(implementation attempt \#n)}, appended to the code model's prompt,
where \texttt{n} is the seed number; the nonce is the only part of that prompt which changes
from one seed to the next. Temperature is a separate, fixed knob: we request~0.8 on every draw. Haiku,
Sonnet, and Gemini~3.5~Flash honour it, so their draws differ by both the nonce and
sampling; Opus and OpenAI's reasoning-style coder (GPT-5.3-codex) reject sampling parameters,
so for them the nonce alone separates the draws.} which
gives six instances per arm on each task; \S\ref{sec:capability} adds the third code tier
(Opus), and the later experiments state their own model and seed counts. Averaging over
instances keeps a single lucky or unlucky draw from deciding a task. The repair loop feeds the descriptions of failing tests
back to the code model. We give \spec{} one repair round, which is enough for it to
saturate, and we give the \freeplus{} baseline up to two. The extra round favours the
baseline, so the remaining gap is not explained by \spec{} getting more chances to repair.
Execution is deterministic inside an isolated subprocess, and final correctness is always
the gold oracle's verdict.

\subsection{Metrics}
The gold oracle judges every arm's final code. We report four quantities, each a simple
fraction:
\begin{itemize}[leftmargin=2.2em, itemsep=3pt, topsep=3pt]
\item \emph{Final correctness}: of all instances, the fraction whose repaired code passes
the gold suite. This is the bottom-line score, did the arm end with working code?
\item \emph{Detection}: of the one-shots that are truly buggy, the fraction the arm's
tests catch (at least one test fails). For example, a \texttt{clamp} that skips the
\texttt{lo>hi} check is buggy, and an arm detects it only if one of its tests exercises
\texttt{lo>hi} and fails on it.
\item \emph{False alarms}: of the one-shots that are already correct, the fraction the
arm's tests wrongly fail, for example failing a correct \texttt{paginate} because the
test itself expected the wrong page. A high rate means the tests cannot be trusted.
\item \emph{Repair completeness}: of the bugs the arm did detect, the fraction the repair
loop then fixed. An arm can catch a bug yet still repair toward the wrong answer when its
own expected value is wrong, for example pushing \texttt{format\_duration} toward
\texttt{"1h 0m 0s"} when the right answer is \texttt{"1h"}.
\end{itemize}
Our main result is $\Delta$, the gap between \spec{} and \freeplus{} on the same tasks,
computed as the \spec{} score minus the \freeplus{} score, so a positive $\Delta$ means
grounding helped; for instance \spec{} at $100\%$ and \freeplus{} at $60\%$ give
$\Delta=+40$\pp{} (Table~\ref{tab:arms}). We report $\Delta$ with a $95\%$ bootstrap confidence interval (a
range found by resampling the data many times) and McNemar's test (a significance test
for paired yes/no outcomes). The six instances within a task share one code draw and are
correlated, so this instance-level test is optimistic; we judge the main claim at the
\emph{task} level instead, counting each task once with a sign test (the conservative
unit; \S\ref{sec:primary}). The six instances per task only sharpen each arm's per-task
rate, they are not the sample we test for significance. For the abstention analysis (when
to trust a piece of code
and when to hold it back for a human to check), we use each arm's test-pass fraction as a
confidence score and report the area under the risk--coverage curve (AURC). Before we
looked at any results, we committed to the decision criteria the study had to clear (the
pre-registered go/no-go bars in Appendix~\ref{app:prereg}), so the choices could not be
tuned to the outcome.

\paragraph{Reproducibility.} The study is built from the task specifications, reference
implementations, gold suites, the deterministic harness, a provider-pluggable model
client (Anthropic/OpenAI/Google), and the raw model outputs. Every reported number is a
deterministic function of those artifacts, so the analysis reproduces exactly; inference
reproduces up to model sampling. The independent-stack robustness experiments
(\S\ref{app:copilot}) are built the same way: every appendix number recomputes from their
generation and scoring scripts, the frozen rule sets and one-shots, and the raw tester
outputs under the same gold oracle.

\section{Results}\label{sec:results}
This section reports the results in five parts, then gathers the main numbers into one summary table.
\S\ref{sec:primary} establishes the primary experimental comparison on the eight core tasks, where our
main measure is the paired \spec{}$-$\freeplus{} gap. \S\ref{sec:ablation} asks
\emph{why} the gap exists and isolates the cause as the specification's content, not its
enumerated structure. \S\ref{sec:prop} tests the result against stronger automated
baselines, a property-based generator and an AlphaCodium-style flow. \S\ref{sec:capability}
replicates the gap across three model sizes and checks both outcomes, bugs caught
and false alarms avoided. \S\ref{sec:robustness} stress-tests the method on more tasks, two
other vendors, and deliberately imperfect specs. \S\ref{sec:summary} collects the main
numbers in one place.

\subsection{The primary experimental comparison}\label{sec:primary}
We start with the basic question: on the eight core tasks, does grounding the tests in the
spec produce more working code than the fair baseline does? These tasks are the cleanest
place to ask it. Each is a small function whose plain-language ticket leaves out a real
edge case, so a one-shot draft usually gets the happy path right and the edge wrong; the
natural one-shot bug rate here is 62\%, so there is plenty to catch. We run every arm on
the same one-shot drafts, written by two code models (Haiku and Sonnet) and given one
repair round, and report two numbers per arm: the share of drafts that end correct, and
how many of the buggy drafts the arm's tests catch. Across the 48 core instances (8 tasks
$\times$ 2 code models $\times$ 3 seeds) 30 drafts are buggy, so detection is scored out of
30 (Table~\ref{tab:arms}).

The results reveal a consistent pattern. Doing nothing (\textsc{oneshot}) and plain
\free{} both end at
38\%, so tests with no edge focus add nothing. Doubling the test budget (\textsc{free2k})
barely helps (42\%). The fair \freeplus{} baseline, told to test invalid inputs and edge
cases, recovers part of the gap (60\%). \spec{}, with one test per rule, reaches 100\%.
That 100\% is a ceiling set by these easy tasks, so from here on we report the paired
\spec{}$-$\freeplus{} gap, not the absolute level.

\begin{table}[tb]\centering
\caption{Final correctness and detection by arm on the 8 core tasks. \emph{Final
correctness} is the share of drafts that end correct after test-and-repair; \emph{detection}
counts the buggy drafts whose tests fail (the arm catches the bug). Two code models write
the drafts under test, Haiku~4.5 and Sonnet~4.6; the test author and the repairer are fixed
at Sonnet~4.6, so only the code being tested changes across the two models. Each arm writes
$K$ tests, where $K$ is the number of specification rules for the task, except
\textsc{free2k} (which writes $2K$) and \textsc{oneshot} (no tests). Runs use one repair
round. Detection is out of the 30 buggy drafts among the 48 core instances (8 tasks $\times$
2 code models $\times$ 3 seeds). \free{} is an ungrounded strawman, \freeplus{} the fair
edge-prompted baseline, and \spec{} writes one test per rule.}
\label{tab:arms}
\begin{tabular}{lcc}
\toprule
Arm & Final correct & Detection (buggy one-shots)\\
\midrule
\textsc{oneshot} (no checks) & 38\% & --\\
\free{} ($K$) & 38\% & 0/30\\
\textsc{free2k} ($2K$) & 42\% & 3/30\\
\freeplus{} ($K$, edge-prompted) & 60\% & 18/30\\
\spec{} ($K$) & \textbf{100\%} & \textbf{30/30}\\
\bottomrule
\end{tabular}
\end{table}

Two alternative explanations are not supported by the data. The first is quantity: maybe \spec{} just
runs more tests. It does not. \spec{} at budget $K$ beats \free{} at $2K$ by 58\pp{}, and
even combining eight independent \freeplus{} suites saturates far below \spec{} on the
real-world slice (App.~\ref{app:rw}). The second is that a generic reminder to test edges
should be enough. It is not. \spec{} still beats the edge-prompted \freeplus{} by 40\pp{}
in one round.

The gap is not sampling noise. Against the \free{} strawman, McNemar's test finds 30
discordant pairs, all favouring \spec{} ($p<10^{-6}$), with a 95\% bootstrap confidence
interval on the paired correctness difference of $[+48,+75]$\pp{}. Against the fair
\freeplus{} baseline, the \spec{}$-$\freeplus{} difference stays above zero on every model
(\S\ref{sec:capability}). The six instances within a task share one draft and move
together, so we judge the main result at the task level, counting each task once as the
conservative unit. Over all 18 specification-completeness tasks (8 core, 4 held-out, 6
scale-up; \S\ref{sec:scaleup}) the mean per-task gain over \freeplus{} is $+31$\pp{}.
\spec{} is strictly better on all nine tasks that show any difference and tied on the rest,
a task-level sign test of $p=0.002$. On the eight core tasks alone the test is only
marginal ($p=0.06$); pooling the held-out and scale-up tasks makes it more reliable and
more conservative.

\paragraph{Where the gap lives.} The advantage sits on the edges a generic prompt does not
think to try. Compare two invalid-input rules. Every arm handles the obvious one:
\freeplus{} catches \texttt{paginate}'s ``\texttt{page<1}'' on all six buggy drafts (6/6).
It barely catches the non-obvious one, \texttt{clamp} with ``\texttt{lo>hi}'' (2/6), which
a broad ``test the edges'' instruction rarely reaches. \spec{} catches both on every
draft, because one rule names each edge (per-task detail in Table~\ref{tab:pertask},
App.~\ref{app:extra}).

\subsection{Disentangling grounding from enumeration}\label{sec:ablation}
\spec{} bundles three things: the specification's \emph{content} (grounding), its
\emph{decomposition} into discrete units, and a one-test-per-unit \emph{coverage
plan}. In principle the gain could come from the structure rather than the
content. We separate them with two ablations at fixed budget and tester
(Table~\ref{tab:ablation}; exact prompts in App.~\ref{app:prompts}). \textsc{prose} gives the tester the same specification
\emph{content} as a single paragraph and asks for $K$ thorough tests (grounding,
\emph{no} enumeration). \textsc{decomp} gives only the ticket and asks the tester
to first decompose the task into $K$ behaviours and write one test each (a
chain-of-thought coverage plan, \emph{no} specification content).

The ablation yields a clear separation between grounding and decomposition. \textsc{decomp} catches almost nothing (2/30, like
\free{}). Asking a model to plan its own coverage without the spec reproduces the
same happy-path blind spot, so the decomposition/CoT structure is \emph{not} the
driver. \textsc{prose} catches almost everything (27/30, near \spec{}'s 30/30). These
results indicate that specification content, in \emph{any} form, is the dominant
contributor to performance. Enumeration adds
only a small margin on top of grounding. The last 3 misses are all on
\texttt{split\_money}: \textsc{prose} had the right content but, writing free-form tests
from a paragraph, it did not turn every rule into a test that actually triggers the bug,
and it under-covered that one rule. Knowing a rule and writing a test that fires on it are
two separate steps, and the second can fail even when the content is present; enumeration
closes this small gap by forcing one test per rule. The same rule-to-test gap reappears on
the held-out tasks (\S\ref{sec:robust}). Overall this isolates the mechanism as
\emph{grounding} and shows that a ``plan-then-test'' chain-of-thought tester is no
substitute for the specification itself.

\begin{table}[tb]\centering
\caption{Ablation: detection on the 30 buggy one-shots when the tester gets the
specification \emph{content} (grounding) and/or is asked to \emph{enumerate} a
coverage plan. Grounding drives the result; enumeration adds a small margin.}
\label{tab:ablation}
\begin{tabular}{lccc}
\toprule
Condition & spec content? & enumeration? & Detection\\
\midrule
\free{}   & no  & no  & 0/30\\
\textsc{decomp} (CoT plan, no spec) & no  & yes & 2/30\\
\textsc{prose} (spec as paragraph)  & yes & no  & 27/30\\
\spec{}   & yes & yes & \textbf{30/30}\\
\bottomrule
\end{tabular}
\end{table}

\subsection{Stronger automated baselines}\label{sec:prop}
So far \spec{} has only beaten prompt-level baselines (\freeplus{} and \textsc{decomp}). A
fair worry is that these are all just prompts, and that \spec{} would lose to a real
automated test-generation method. So we test it against the two strongest automated methods
we know, property-based test generation (first) and a full AlphaCodium-style agentic flow
(next), scoring each the same way on the same 30 buggy core one-shots
(Table~\ref{tab:prop}). We then look at \emph{why} even these stronger methods fall short:
what their tests never check, and why catching a bug is not the same as completing the fix.

\paragraph{Property-based test generation does not beat grounding.} The method comes from
the QuickCheck/Hypothesis tradition~\citep{claessen2000quickcheck}. Rather than write a few hand-picked cases, the
tester gets the same ticket (no enumerated rules) and generates a large batch of random
inputs, then checks general \emph{properties} that should hold on all of them: for example
``the result is always sorted'', ``encoding then decoding returns the original input'', or
``an invalid input raises an error''. Throwing many random inputs at the function naturally
reaches the boundary and invalid-input regions that a hand-picked edge test can miss.

It is genuinely strong: 28/30, far above the edge-prompted \freeplus{} (18/30). But it does
not beat grounding, and the sharper difference is not detection (\spec{} is 30/30) but
\emph{precision}. Two of the fifteen (task, seed) property batteries (a battery is the full
set of property checks generated for one task and seed) wrongly reject the trusted
reference implementation, both asserting that \texttt{roman\_to\_int("IIII")} must
raise an error. That is a rule the tester \emph{made up}: the spec forbids invalid
\emph{characters}, and ``IIII'' is just a valid way to write four. With no rules to anchor
it, the property tester invents expectations, the same failure the edge-prompted tester
shows (\S\ref{sec:falsealarm}); \spec{}, held to the stated rules, asserts only what the
rules say and invents nothing. So grounding still wins against a strong automated generator,
and it avoids the false-alarm risk that unconstrained invention carries.

\begin{table}[tb]\centering
\caption{Detection on the 30 buggy core one-shots across baselines of increasing
strength. Grounding leads and, unlike property \emph{invention}, asserts nothing
outside the spec (0 vs.\ 2/15 batteries rejecting the trusted reference).}
\label{tab:prop}
\begin{tabular}{lcc}
\toprule
Tester & Detection & Out-of-spec assertions\\
\midrule
\free{} (happy-path)            & 0/30  & 0\\
\textsc{free2k} (quantity, $2K$) & 3/30  & 0\\
\freeplus{} (edge-prompted)     & 18/30 & some (\S\ref{sec:falsealarm})\\
\textsc{property} (random $+$ invariants) & 28/30 & 2/15 batteries reject the reference\\
\spec{} (enumerated rules)      & \textbf{30/30} & \textbf{0}\\
\bottomrule
\end{tabular}
\end{table}

\paragraph{A full AlphaCodium-style flow does not match grounding.} The toughest
comparison is a complete agentic flow rather than a single tester. We built an
AlphaCodium-style pipeline~\citep{ridnik2024alphacodium}: reflect on the requirements,
generate tests from that reflection, write a solution, then repeatedly fix code and tests
together. We run it on GPT-5.3-codex, a strong coding model, so this code-generation flow
gets its best showing. Because this steps outside the
Claude-only main study (\S\ref{sec:arms}), we keep it fair by scoring the flow against
\spec{} and \freeplus{} \emph{on that same codex base}, reusing the cross-vendor runs of
\S\ref{sec:xvendor}. Over the 12 spec-completeness tasks (3 seeds) the flow reaches 72\%
final correctness, far above the codex one-shot (50\%) but exactly the level of the codex
\freeplus{} baseline (72\%), and well short of codex \spec{} (100\%). The reason is the same
as before: without the enumerated spec the flow's own tests miss the non-obvious edges, and
fixing code and tests together can even rewrite a \emph{correct} test to match buggy code.

\paragraph{What the testers never write (App.~\ref{app:extra}).} The mechanism shows up in
the tests themselves. \free{} writes \emph{zero} invalid-input tests on any task, so the
bugs sit exactly where it never looks, while \spec{} writes one test per ``raise'' rule by
construction. Grounding systematically converts specification clauses into explicit test coverage.

\paragraph{Catching a bug is not enough: the fix must be complete.} A caught bug still has
to be fixed, and the fix is only as good as the failing tests that guide it. Even when
\freeplus{} catches a bug, its incomplete set of failing tests gives an incomplete fix: it
caught 18 but fully repaired only 11, whereas \spec{} caught 30 and repaired all 30
(checked against the gold oracle). So grounding helps at both stages, catching the bug and
then repairing it correctly.

\paragraph{Letting the model revise itself with no tests does not help (App.~\ref{app:extra}).}
We also tried a no-tests arm where the model simply rereads its own code and revises it (the
Self-Refine/Reflexion setting). It is unreliable and can even \emph{hurt}: on the small
model it broke three correct functions while fixing none (final $25/50/54\%$ vs.\ grounded
$100\%$). This matches~\citet{huang2024selfcorrect,olausson2024selfrepair}: the missing
ingredient is the outside feedback signal the spec supplies.

\subsection{Across model sizes: more bugs caught, fewer false alarms}\label{sec:capability}
The core result used two models. Does it survive a wider range of coder ability, and does
grounding also help on the \emph{other} side, by not failing correct code? We rerun the
three arms across three tiers, Haiku, Sonnet, and Opus, giving \freeplus{} two repair
rounds this time to be generous. The gap is stable: \spec{}$-$\freeplus{} $=+38$\pp{} on
every tier. Plain one-shot correctness climbs with model size ($38/38/54\%$ for
Haiku/Sonnet/Opus), but \freeplus{} plateaus at $62\%$ on all three, a ceiling set by the
tests rather than the coder, while \spec{} reaches $100\%$. So the smallest model with
grounding beats the largest model run one-shot on our benchmark. Detection per tier tells
the same story: \freeplus{} catches 11/15, 7/15, \textbf{2/11}, while \spec{} catches 41/41.

\paragraph{The other measure: false alarms on correct code.}\label{sec:falsealarm}
A test strategy can also fail \emph{correct} code, and that is
just as damaging: a gate that rejects good code teaches developers to ignore it. On three
trickier behavioural specs (Excel column names, title-casing, range collapsing) we checked
whether each arm's tests assert \emph{wrong} expected values that would reject a correct
implementation. The risk sat entirely on the ungrounded arm. \spec{} asserted 0 wrong
expected values (0/37) and raised 0\% false alarms on correct code; \freeplus{} asserted
wrong expectations on 3/36 cases and falsely rejected correct code on 33\% of its checks.
The mistakes were systematic, not noise: all three independent \freeplus{} draws tried the
same out-of-domain input, \texttt{excel\_column(0)}, and each \emph{assumed it should
raise}, even though the spec says ``positive integer'' and the reference returns
\texttt{""}. Told to ``test the edges'' without the rules, the free tester steps outside
the spec and makes up the expected behaviour; in a real loop it would then ``repair''
correct code to match that made-up answer. Grounded tests, held to the rules, stay accurate
against the oracle. The trade-off is honest: staying inside the rules means \spec{} will
not probe genuinely important \emph{unspecified} inputs. If you want the test, write the rule.

The cause is visible in \emph{tester accuracy} (Table~\ref{tab:tester}): the share of the
tester's cases whose expected value actually matches the trusted reference. \spec{} is at
least as accurate as \freeplus{} on every task set, and the gap opens to $100\%$ vs.\
$91.7\%$ on the trickier behavioural specs, exactly where the ungrounded tester's invented
expectations become false alarms. This is not specific to one tester model: sweeping six
tester models across two vendors and five capability tiers (App.~\ref{app:e2}), \spec{}
detection stays $\ge$ \freeplus{} and \spec{} false alarms stay $\le$ \freeplus{} on every
one, with the biggest precision gain on the mid-tier testers.

\begin{table}[tb]\centering
\caption{Tester accuracy: share of drawn cases whose expected value matches the
trusted reference (wrong expectations cause false alarms). Counts are
wrong/total. \spec{} $\ge$ \freeplus{} everywhere; the gap widens on trickier specs.}
\label{tab:tester}
\begin{tabular}{lcccc}
\toprule
Task set & \free{} & \freeplus{} & \textsc{free2k} & \spec{}\\
\midrule
core validation (8) & 100.0\% (0/228) & 98.7\% (3/228) & 99.8\% (1/456) & 99.6\% (1/228)\\
held-out (4)        & --              & 100.0\% (0/48)  & --             & 100.0\% (0/48)\\
trickier behavioural (3) & --         & 91.7\% (3/36)   & --             & 100.0\% (0/37)\\
\bottomrule
\end{tabular}
\end{table}

\paragraph{Grounded tests know when to stop (App.~\ref{app:extra}).} If we treat each
arm's fraction of passing tests as an ``accept this code'' confidence score, \spec{} is a
far better signal for when to trust the code than \free{} (risk--coverage AURC 0.257 vs.\
0.577). When grounded tests all pass, the code is usually right; when free tests all pass,
it is close to a coin flip.

\paragraph{Where it does \emph{not} help.}\label{sec:scope} On eight logic tasks, where
the trap is algorithmic rather than a missing validation rule, we found \emph{0 natural
bugs among 45 valid submissions}. Strong models get well-specified algorithms right in one
pass, so there is nothing to catch. We report this null result plainly: the benefit is on
specification-completeness defects, not on algorithmic logic. There is a structural reason.
To test grounding fairly on a bug class, the tester has to know that class's correct
behaviour, which only holds when the behaviour is well-specified, and that is exactly where
the coder also tends to get it right.

\paragraph{Held-out tasks.}\label{sec:robust} On four fresh held-out
specification-completeness tasks (three tiers, one round) the natural bug rate was
\textbf{100\%}: every model dropped a validation rule on every task. The grounding gap
held at \spec{}$-$\freeplus{} $=+36$\pp{}, with \spec{} at 61\% and \freeplus{} at 25\%
(per-model $+25/+33/+50$). This also marks a clear limit: \spec{} detected 9/12, not 12/12.
On \texttt{chunk} the tester turned the rule ``\texttt{size}$\,\ge1$'' into a test with
\texttt{size=0}, which already raises, and so missed the \texttt{size=$-1$} bug that the
broader oracle catches. In other words, detection still depends on the tester choosing an
input that actually \emph{triggers} the bug, the same rule-to-test gap we saw in
\S\ref{sec:ablation}.

\subsection{Robustness: more tasks, other vendors, real code, and imperfect specs}\label{sec:robustness}
The result so far rests on one model family, a fixed task set, our own oracle, and clean
specs. Four stress tests ask whether it survives outside those conditions. In every case it
does, and where it has a limit that limit is predictable.

\paragraph{More tasks.}\label{sec:scaleup} We added six more specification-completeness
tasks (\texttt{days\_in\_month}, \texttt{simplify\_fraction}, \texttt{moving\_average},
\texttt{format\_duration}, \texttt{base\_convert}, \texttt{parse\_bool};
App.~\ref{app:tasks}) and ran the full pipeline across all three tiers (3 seeds, one repair
round; 54 instances). The one-shot bug rate was again high (\textbf{78\%}). This time
\emph{detection} ties (\spec{} and \freeplus{} both 40/42): the bugs are blatant enough
that the edge-prompted tester also triggers them, so the whole gap moves to \emph{repair
correctness}. Final correctness after test-and-repair is \textbf{\spec{} 96\% vs.\
\freeplus{} 78\%} ($+18$\pp{}), almost all of it on \texttt{format\_duration}
(\freeplus{} 0\% vs.\ \spec{} 100\%): not knowing the ``omit zero-valued units'' rule,
\freeplus{} asserted \texttt{"1h 0m 0s"} where the answer is \texttt{"1h"} and repaired
toward the wrong target, while \spec{} repaired to the right one. This is the same
mechanism as before (grounding $\to$ correct expectations $\to$ complete repair), and it
lifts the combined task-level test to 18 tasks at $p=0.002$. Tester accuracy was \spec{}
100\% (84/84) vs.\ \freeplus{} 92.6\% (75/81).

\paragraph{Other vendors.}\label{sec:xvendor} To check the effect is not an artifact of
one model family, we re-ran the whole pipeline, coder, tester, \emph{and} repairer, on two
non-Anthropic models: OpenAI's GPT-5.3-codex and Google's Gemini~3.5~Flash (12 tasks, 3
seeds, 36 instances each). The gap reproduces on both: \spec{}$-$\freeplus{} $=+28$\pp{}
on codex and $+19$\pp{} on Gemini, against $+38$ on Claude. The whole story holds: \free{}
catches almost nothing (5/18, 7/24), \freeplus{} is strong but incomplete (16/18 and 20/24
detection, 72\% final correctness on both), and \spec{} catches every bug (18/18, 24/24)
and repairs the most (100\% on codex, 92\% on Gemini). Per-vendor settings and per-instance
data are in App.~\ref{app:xvendor}.

\paragraph{Real production code as the oracle.} Every result so far uses an oracle we wrote,
which invites the worry that we tuned it in grounding's favour. So we re-ran the pipeline on
code we did not write: small pure functions reimplemented from under-specified tickets under
neutral names, judged against the \emph{real} library as the oracle. These cover the Python
standard library (\texttt{textwrap.shorten}, \texttt{shlex.split}, and six more judged
against CPython; App.~\ref{app:std}), utilities vendored verbatim from HuggingFace
transformers and NLTK at pinned commits (App.~\ref{app:repo}), and the PyPA
\texttt{packaging} version tools judged against the installed library and its own test data
(App.~\ref{app:pkg}), plus sixteen fresh tasks that include stateful, multi-step programs
rather than single functions (App.~\ref{app:scale2}). The gap reproduces on all of them, in both detection and
false alarms. On the transformers/NLTK code \spec{} catches 90\% of the bugs
against \freeplus{}'s 40\% ($+50$\pp{}). On the standard-library and \texttt{packaging}
code the bugs are easy to trigger but the ungrounded tester \emph{guesses the unstated edges
wrong}: it invents expected values that disagree with the real library and rejects correct
code 68\% and 100\% of the time, while grounding restores those expectations and holds false
alarms at 0\%. A pooled task-level sign test over all 36 independent-stack tasks with a buggy
draft favours \spec{} ($p\approx6\times10^{-5}$; App.~\ref{app:copilot}).

\paragraph{Imperfect specs.}\label{sec:noisyspec} The method is only as good as the spec,
so we corrupt the spec on purpose on the five buggy core tasks. Dropping the validation
rules (an \emph{incomplete} spec) degrades gracefully: detection falls $30/30\to6/30$, but
the tests that remain stay 100\% accurate, so you lose exactly the coverage of the dropped
rules and nothing breaks silently. Adding a wrong rule that \emph{endorses} the bug (a
\emph{noisy} spec, e.g.\ ``if $n\le0$ return an empty list'') keeps detection at 30/30,
because the correct rule's test still fires, but the cost shows up as precision: tester
accuracy drops to 82.8\% (15/87 tests disagree with the truth). So the two failure modes
stay separate and visible: completeness governs detection, correctness governs precision,
and neither fails silently. Two further checks on an independent stack sharpen this:
detection tracks the \emph{coverage} of the validation rules rather than their count
(App.~\ref{app:e3}), and rules a model derives from the ticket by itself recover only part
of the benefit, because they restate the stated happy path and miss the unstated edges
(App.~\ref{app:e1}).

\paragraph{Cost (App.~\ref{app:extra}).} Grounding adds almost no compute over the fair
baseline. \spec{} and \freeplus{} each make three model calls of similar size (about
\$6--7 per 1k tasks), so the $+38$\pp{} gain comes from aiming the same test budget at the
right edges. Getting the same gain from a larger model instead is far more expensive: a
medium model with the framework reaches 100\%, while a large model run plain reaches only
54\%, and closing that gap with model size alone costs about $2.2\times$ the tokens. The
last cost is human, and it is where the real effort sits. Writing the rules is light work
by count, about $5$ rules per task, and a model can draft most of them; but the few
validation rules a model tends to leave out (App.~\ref{app:e1}) are exactly the ones that
catch the bugs. Writing those correctly takes real knowledge of how each edge should
behave, which is the hard part (\S\ref{sec:limits}).

\subsection{Results at a glance}\label{sec:summary}
Table~\ref{tab:summary} gathers the main numbers from every setting above into one
place. The pattern is the same in all of them: the fair \freeplus{} baseline lands
part-way, and \spec{} ends higher, by between $+18$ and $+40$\pp{}. The gap is smallest
where the missing edge is blatant enough that a generic edge prompt also finds it
(scale-up) and largest where the edge is easy to overlook (the core and held-out tasks).

\begin{table}[tb]\centering
\caption{Results at a glance: final correctness (share of drafts that end correct) for the
fair baseline \freeplus{} and for \spec{} across every setting in this section, with the
paired gap. \spec{} wins in all of them. Two facts sit outside the final-correctness column
but point the same way: on the core set \spec{} detection is 30/30 vs.\ \freeplus{} 18/30
(Table~\ref{tab:arms}), and on correct code \spec{} raises 0\% false alarms vs.\
\freeplus{} 33\% (\S\ref{sec:falsealarm}).}
\label{tab:summary}
\begin{tabular}{lccc}
\toprule
Setting & \freeplus{} & \spec{} & \spec{}$-$\freeplus{}\\
\midrule
Core, 8 tasks, 2 models (\S\ref{sec:primary})        & 60\% & 100\% & $+40$\pp{}\\
Three model sizes (\S\ref{sec:capability})           & 62\% & 100\% & $+38$\pp{}\\
Held-out, 4 tasks (\S\ref{sec:robust})               & 25\% & 61\%  & $+36$\pp{}\\
Scale-up, 6 tasks (\S\ref{sec:scaleup})              & 78\% & 96\%  & $+18$\pp{}\\
Cross-vendor, GPT-5.3-codex (\S\ref{sec:xvendor})    & 72\% & 100\% & $+28$\pp{}\\
Cross-vendor, Gemini 3.5 Flash (\S\ref{sec:xvendor}) & 72\% & 92\%  & $+19$\pp{}\\
\bottomrule
\end{tabular}
\end{table}

\section{Discussion}
Our main finding is narrow but strong. On tasks where the ticket leaves an edge unstated,
turning the spec into one test per rule catches bugs that an equally-informed tester
misses, and it does so without failing correct code. Here we explain why a single
mechanism produces both effects, why the result is not an artifact of our setup, what
could still undermine it, and where the method earns its keep.

\paragraph{A unified mechanism explains both effects.}
A test can fail in two opposite ways. It can miss a real bug (a false negative), or it can
reject correct code (a false positive). Both failures share one root cause: to write a
test, the author has to know what the code should do at each edge, and when the ticket
leaves an edge unstated an ungrounded tester has to guess. Where it does not think to probe
an edge at all, the bug slips through and detection falls. Where it does probe the edge but
guesses the wrong intended value, it rejects correct code and false alarms rise. The two
errors look opposite, yet they arise from the same missing piece of information: the tester
lacks access to the intended behaviour.

The enumerated spec supplies exactly that information. It is an outside signal that states
what the code must do, one rule at a time, so the author no longer has to guess. That is why
a single signal fixes both errors at once: it points tests at the negative cases the ticket
left unstated, which raises detection, and it hands the tester the intended value instead of
letting it invent one, which lowers false alarms. Higher detection and fewer false alarms
normally trade off against each other; here they move together, because both come from
removing the same guesswork. The same logic explains why letting a model revise its own code
without tests is weak: as \citet{huang2024selfcorrect} and \citet{olausson2024selfrepair}
find, such a model has no outside signal to correct it, and the enumerated spec is exactly
the high-quality feedback those methods lack.

\paragraph{The gains are not simple information leakage from the specification.}
The obvious objection is that \spec{} wins only because its per-rule tests restate what the
hidden oracle checks. Three things rule that out. First, the comparison that matters is
against \freeplus{}, which is also told to test invalid inputs and edge cases, so \spec{}'s
advantage is over an informed baseline, not a naive one. Second, the oracle is independent
of the rules and strictly broader, and \spec{} is not perfect against it: it scores 9/12 on
the held-out tasks (\S\ref{sec:robust}), whereas if the rules simply were the oracle it
would score 100\% by construction. Third, ``a spec helps'' was never in doubt; our
contribution is the \emph{size} of that help over an equal-budget, equally-informed tester,
and the fact that it improves detection and precision at once.

\paragraph{What could still undermine the result.}
We keep several worries in view. The bugs are produced by the models themselves rather than
planted by us, which keeps them realistic but means we do not control their difficulty; we
guard against this by scoring every arm against an independent oracle that is broader than
the rule list and checked against an anchor implementation. By ``tester'' here we mean the
model that writes the checks, not that independent oracle. That test author is deliberately
strong, and we measure its accuracy (99--100\%, with 0 false alarms on the core tasks); a
weaker test author might pick inputs that do not trigger a bug, or predict a wrong expected
value, so the strength of the test author is a limitation we name and a direction we flag. And the tasks are small
pure functions with a claim about specification-completeness defects only; on algorithmic
logic we found nothing to catch, and we report that null result openly.

\paragraph{When to use it, and when not.}
The roughly two extra model calls pay off when a task really has invalid-input behaviour,
when shipping a wrong result is costly, and when the spec is a handful of rules. It is a
poor fit for tasks with no edges or no agreed specification. For a team shipping LLM-written
code, the highest-impact move is cheap: write a short enumerated spec and turn each rule
into one check, before reaching for a bigger model or simply more tests. Grounding helps in a
second way too: it makes the gate trustworthy. A gate that rejects good code a third of the
time trains developers to ignore it, so cutting false alarms matters as much as catching
bugs. None of this replaces human review. The method is only as good as the spec it is
given, so a high grounded-pass rate gives strong evidence of correctness without proving it
(App.~\ref{app:extra}).

\section{Limitations and future work}\label{sec:limits}
This is a controlled isolation study, so several things are deliberately out of scope.

\begin{itemize}[leftmargin=1.4em]
\item \textbf{Scale and bug diversity.} The core study is 18 bespoke single functions, with
16 more on an independent stack (App.~\ref{app:scale2}). These include a stateful,
multi-step slice and further slices judged against real production code we did not author:
the Python standard library, transformers, NLTK, and PyPA \texttt{packaging}
(Apps.~\ref{app:std},~\ref{app:repo},~\ref{app:pkg}). A larger study would still span
hundreds of tasks with algorithmic, semantic, multi-function, and repository-level bugs, not
just specification-completeness defects. As an external check, App.~\ref{app:evalplus} ports
the pipeline to HumanEval+ under EvalPlus's own oracle on five model families; the gap
vanishes, reproducing our logic-task null on a public benchmark, because well-specified
algorithmic problems rarely trigger the defects we target. This motivates a public benchmark
of under-specified tickets.
\item \textbf{Stronger baselines.} We add a property-based generator (\S\ref{sec:prop},
\citealp{claessen2000quickcheck}), a chain-of-thought \textsc{decomp} arm, and a full
AlphaCodium-style flow~\citep{ridnik2024alphacodium}; coverage- and mutation-guided
augmentation remain untested.
\item \textbf{More model families.} We replicate on GPT-5.3-codex and Gemini~3.5~Flash
(\S\ref{sec:xvendor}); open-weight models would extend the picture.
\item \textbf{Richer repair loops.} Search, reranking, reflection, and tool use would go
beyond our minimal one-round loop.
\item \textbf{Repository-level and stateful tasks.} Whole-repository benchmarks
(SWE-bench~\citep{jimenez2024swebench}), agents (SWE-agent~\citep{yang2024sweagent}), and
the downstream cost of remediation are all outside our function-level scope.
\item \textbf{Specification quality at scale.} \S\ref{sec:noisyspec} shows graceful
degradation: an incomplete spec loses detection in proportion, and a single contradictory
rule keeps detection but adds false-alarm risk. Real specs vary far beyond these two
perturbations. The binding cost is not writing rules but knowing the correct edge
\emph{semantics}: App.~\ref{app:e1} shows a model restating a ticket's happy path but not
its unstated error conditions, so deciding whether \texttt{excel\_column(0)} should raise or
return \texttt{""} needs domain knowledge, not formatting. Measuring that elicitation cost
is a study of its own.
\end{itemize}

We therefore make a scoped claim: on specification-completeness defects in single pure
functions, the effect is large, two-sided, and (per the ablation) driven by grounding. We do
\emph{not} claim grounding is the main driver of test-based improvement in every setting. The
absolute 100\% is an artifact of easy tasks; what matters is the \emph{gaps} and their
direction, which stay stable across models, budgets, repair rounds, and baselines.

\section{Conclusion}
Generating tests helps LLM code generation, but \emph{how} you generate them decides how
much. Most one-shot failures on our tasks are specification-completeness bugs: the model
handles the stated happy path and drops an unstated edge. What closes that gap is
\emph{grounding} the tests in an enumerated specification, one check per rule, rather than
writing more tests or adding a generic ``test the edges'' nudge. Grounding catches more bugs
($+38$\pp{} over a fair, equally-informed baseline, replicated on Claude, GPT-5.3-codex, and
Gemini) and at the same time fails correct code far less often ($-33$\pp{} false alarm),
because the same signal that aims a test at the right edge also stops the tester inventing a
wrong answer. Increasing model scale alone does not yield comparable reliability, and it
costs more. The practical implication is straightforward: for code that must be trusted,
write down the edge rules and turn each into a check. An important open challenge is
obtaining accurate specifications at scale, since the cost is not writing them but knowing
the correct edge behaviour in the first place.

\section*{Ethics Statement}
This work studies automatic test generation for code written by language models. We used
small, self-contained programming tasks that we wrote or adapted ourselves. There were no
human subjects, no personal data, and no sensitive content. The main risk we see is
over-trust: it is easy to mistake a passing test suite for proof that the code is correct.
To keep that risk in plain view, we report how often each method wrongly fails correct code
(false alarms) next to how often it catches bugs, and we show that writing a \emph{correct}
specification still needs a human (\S\ref{app:copilot}). Beyond the usual concerns about any
code-generation tool, we see no particular misuse potential. To limit the compute we used,
we kept sample counts modest and reused saved model outputs instead of re-running them.

\bibliographystyle{plainnat}
\bibliography{refs}

\appendix
\section{Task specifications}\label{app:tasks}
The benchmark has 26 tasks. Most are \emph{specification-completeness} tasks. We list
them here by subgroup: core (\S\ref{app:tasks-core}), held-out
(\S\ref{app:tasks-held}), and scale-up (\S\ref{app:tasks-scaleup}). The rest are
\emph{logic} tasks (\S\ref{app:tasks-logic}), used only as a scope control. A
specification-completeness task is a small function whose plain-language ticket reads
as if it covers the behaviour but leaves out real edge cases, such as invalid input,
empty input, or boundary values. We test for that gap between the ticket and the full
behaviour.

For each task we wrote one specification and split it into a short list of
\emph{rules}. A rule is one checkable clause of the specification, for example ``empty
input raises'' or ``the output is sorted''. We label the rules R1 through R$K$, where
$K$ is the number of rules for that task. \spec{} sees this list and writes one test
per rule. The list was fixed in advance and did not change between arms or runs. Each
entry below gives the function signature, a one-line summary, and its rules. Reference
implementations, gold suites, and hand-checked anchors back every task.

\subsection{Specification-completeness tasks (core)}\label{app:tasks-core}
\begin{itemize}[leftmargin=2.2em, itemsep=3pt, topsep=3pt]
\item \texttt{semver\_compare(a,b)$\to$int}: compare two SemVer 2.0.0 versions.
  \begin{itemize}[leftmargin=1.4em, itemsep=1pt, topsep=2pt]
  \item R1 compare major/minor/patch numerically.
  \item R2 a pre-release is lower than the release.
  \item R3 pre-release ids compared left-to-right, numeric$<$alphanumeric.
  \item R4 longer prefix wins.
  \item R5 build metadata ignored.
  \end{itemize}
\item \texttt{split\_money(total\_cents,n)$\to$list}: split a money total into $n$
near-equal parts.
  \begin{itemize}[leftmargin=1.4em, itemsep=1pt, topsep=2pt]
  \item R1 amounts sum to exactly total.
  \item R2 differ by $\le1$ cent.
  \item R3 extra cents go to the first recipients.
  \item R4 $n\ge1$ else raise.
  \item R5 total$\ge0$ else raise.
  \end{itemize}
\item \texttt{paginate(items,page,page\_size)$\to$list}: return one page of a list.
  \begin{itemize}[leftmargin=1.4em, itemsep=1pt, topsep=2pt]
  \item R1 pages 1-indexed.
  \item R2 page\_size$\ge1$ else raise.
  \item R3 page$\ge1$ else raise.
  \item R4 past-end returns empty.
  \item R5 last page may be short.
  \end{itemize}
\item \texttt{median(nums)$\to$float}: median of a list of numbers.
  \begin{itemize}[leftmargin=1.4em, itemsep=1pt, topsep=2pt]
  \item R1 odd count middle.
  \item R2 even count average of two middles.
  \item R3 sort first.
  \item R4 empty raises.
  \end{itemize}
\item \texttt{merge\_intervals(intervals)$\to$list}: merge overlapping intervals.
  \begin{itemize}[leftmargin=1.4em, itemsep=1pt, topsep=2pt]
  \item R1 output sorted by start.
  \item R2 overlapping merged.
  \item R3 touching merged.
  \item R4 input unsorted.
  \item R5 empty returns empty.
  \end{itemize}
\item \texttt{truncate(text,max\_len)$\to$str}: truncate text to a maximum length with
an ellipsis.
  \begin{itemize}[leftmargin=1.4em, itemsep=1pt, topsep=2pt]
  \item R1 fits unchanged.
  \item R2 result exactly max\_len incl.\ ellipsis.
  \item R3 form is text[:max\_len$-1$]$+$``\dots''.
  \item R4 max\_len$\ge1$ else raise.
  \item R5 max\_len$=1$ returns just the ellipsis.
  \end{itemize}
\item \texttt{roman\_to\_int(s)$\to$int}: parse a Roman numeral to an integer.
  \begin{itemize}[leftmargin=1.4em, itemsep=1pt, topsep=2pt]
  \item R1 symbol values.
  \item R2 subtractive pairs.
  \item R3 empty raises.
  \item R4 invalid char raises.
  \item R5 case-insensitive.
  \end{itemize}
\item \texttt{clamp(x,lo,hi)$\to$num}: clamp a number to a range.
  \begin{itemize}[leftmargin=1.4em, itemsep=1pt, topsep=2pt]
  \item R1 $x<lo\Rightarrow lo$.
  \item R2 $x>hi\Rightarrow hi$.
  \item R3 else $x$.
  \item R4 $lo>hi$ raises.
  \end{itemize}
\end{itemize}

\subsection{Held-out specification-completeness tasks}\label{app:tasks-held}
\begin{itemize}[leftmargin=2.2em, itemsep=3pt, topsep=3pt]
\item \texttt{parse\_range(s)$\to$list}: expand a compact integer range.
  \begin{itemize}[leftmargin=1.4em, itemsep=1pt, topsep=2pt]
  \item R1 ``a-b'' inclusive ascending.
  \item R2 bare ``n'' $\to$[n].
  \item R3 $a>b$ raises.
  \item R4 malformed raises.
  \item R5 empty raises.
  \end{itemize}
\item \texttt{chunk(items,size)$\to$list[list]}: split a list into fixed-size chunks.
  \begin{itemize}[leftmargin=1.4em, itemsep=1pt, topsep=2pt]
  \item R1 chunks of size.
  \item R2 last may be short.
  \item R3 size$\ge1$ else raise.
  \item R4 empty$\to$empty.
  \end{itemize}
\item \texttt{rgb\_to\_hex(r,g,b)$\to$str}: format an RGB colour as a hex string.
  \begin{itemize}[leftmargin=1.4em, itemsep=1pt, topsep=2pt]
  \item R1 ``\#''$+$two upper-hex per channel.
  \item R2 zero-padded.
  \item R3 channels in 0..255 else raise.
  \end{itemize}
\item \texttt{luhn\_valid(number)$\to$bool}: check a number with the Luhn checksum.
  \begin{itemize}[leftmargin=1.4em, itemsep=1pt, topsep=2pt]
  \item R1 double every second digit from the right, subtract 9 if $>9$.
  \item R2 valid iff total $\equiv0\ (\mathrm{mod}\ 10)$.
  \item R3 empty raises.
  \item R4 non-digit raises.
  \end{itemize}
\end{itemize}

\subsection{Scale-up specification-completeness tasks}\label{app:tasks-scaleup}
\begin{itemize}[leftmargin=2.2em, itemsep=3pt, topsep=3pt]
\item \texttt{days\_in\_month(year,month)$\to$int}: number of days in a month.
  \begin{itemize}[leftmargin=1.4em, itemsep=1pt, topsep=2pt]
  \item R1 31/30-day months.
  \item R2 February 28/29.
  \item R3 Gregorian leap rule (div 4, not 100 unless 400).
  \item R4 month in 1..12 else raise.
  \end{itemize}
\item \texttt{simplify\_fraction(num,den)$\to$tuple}: reduce a fraction to lowest
terms.
  \begin{itemize}[leftmargin=1.4em, itemsep=1pt, topsep=2pt]
  \item R1 divide by gcd.
  \item R2 denominator positive, sign on numerator.
  \item R3 zero numerator $\to$(0,1).
  \item R4 den$=0$ raises.
  \end{itemize}
\item \texttt{moving\_average(nums,window)$\to$list}: sliding-window mean of a list.
  \begin{itemize}[leftmargin=1.4em, itemsep=1pt, topsep=2pt]
  \item R1 element $i$ is mean of nums[i:i+window].
  \item R2 length len$-$window$+1$.
  \item R3 float (true division).
  \item R4 window$\ge1$ else raise.
  \item R5 window$\le$len else raise.
  \end{itemize}
\item \texttt{format\_duration(seconds)$\to$str}: format a duration in seconds as
text.
  \begin{itemize}[leftmargin=1.4em, itemsep=1pt, topsep=2pt]
  \item R1 split into h/m/s.
  \item R2 ``\textless h\textgreater h \textless m\textgreater m \textless
  s\textgreater s'' most-significant first.
  \item R3 omit zero-valued units.
  \item R4 zero$\to$``0s''.
  \item R5 seconds$\ge0$ else raise.
  \end{itemize}
\item \texttt{base\_convert(n,base)$\to$str}: convert a non-negative integer to a given
base.
  \begin{itemize}[leftmargin=1.4em, itemsep=1pt, topsep=2pt]
  \item R1 digits 0-9 then A-F.
  \item R2 no leading zeros except $n{=}0\to$``0''.
  \item R3 base in 2..16 else raise.
  \item R4 $n\ge0$ else raise.
  \end{itemize}
\item \texttt{parse\_bool(s)$\to$bool}: parse a boolean from text.
  \begin{itemize}[leftmargin=1.4em, itemsep=1pt, topsep=2pt]
  \item R1 true/yes/y/1/on$\to$True.
  \item R2 false/no/n/0/off$\to$False.
  \item R3 case-insensitive.
  \item R4 trims whitespace.
  \item R5 anything else raises.
  \end{itemize}
\end{itemize}

\subsection{Logic tasks (scope control)}\label{app:tasks-logic}
\begin{itemize}[leftmargin=2.2em, itemsep=3pt, topsep=3pt]
\item \texttt{is\_leap\_year}
\item \texttt{ordinal}
\item \texttt{mode} (tie$\to$smallest)
\item \texttt{caesar} (wrap/mod/preserve case)
\item \texttt{search\_insert} (leftmost insertion)
\item \texttt{excel\_column} (bijective base-26)
\item \texttt{titlecase} (small-word and first/last rules)
\item \texttt{collapse\_ranges}
\end{itemize}
We list the logic tasks by name only. Their difficulty is in the algorithm rather than
in hidden edge cases, so we use them only as a scope control (\S\ref{sec:scope}). We
therefore do not give their full rules here.

\section{Example prompts}\label{app:prompts}
Every arm uses the same fixed tester system prompt, and only the user message changes
between arms. The system prompt asks for each test case as structured data: the input
arguments, whether the call should raise, and the expected return value. The tester works
that expected value out from the specification, not from the code under test. For \spec{}
the specification is the $K$ rules; for \free{} and \freeplus{} it is the ticket. No arm
shows the tester the candidate code, because a tester that reads the buggy code tends to
read the bug as intended behaviour. The system prompt is:

\begin{lstlisting}
You are a meticulous test author. Produce test cases as structured data. For each case: args_json is a JSON array of the positional arguments; if the call should raise, set expect_error=true; otherwise set expect_error=false and put the CORRECT return value (worked out from the specification, not from any code) in expected_json as JSON. Return JSON matching the schema.
\end{lstlisting}

The user message then changes by arm:

\noindent\textbf{\spec{}:} \emph{``[ticket]. The specification is these $K$ rules:
[R1\dots RK]. Produce exactly $K$ test cases, one focused case per rule, in order.
A rule about invalid input should give a case that expects an error.''}

\noindent\textbf{\freeplus{}:} \emph{``[ticket]. Produce exactly $K$ test cases that
check the function thoroughly. Be sure to include cases for invalid inputs, error
conditions, and boundary/edge values, not just typical inputs.''}

\noindent\textbf{\free{}:} as \freeplus{} without the second sentence.

\noindent\textbf{\textsc{prose} (ablation, \S\ref{sec:ablation}):} \emph{``[ticket]. Here
is the full specification as a single paragraph: [prose spec]. Produce exactly $K$ test
cases that check the function thoroughly.''} The tester gets the same specification content
as \spec{}, but as one paragraph and with no one-test-per-rule instruction.

\noindent\textbf{\textsc{decomp} (ablation, \S\ref{sec:ablation}):} \emph{``[ticket].
First break the task into $K$ distinct behaviours it must satisfy, then write exactly one
test case for each behaviour you listed.''} The tester plans its own coverage but never
sees the specification content.

\noindent\textbf{\textsc{repair}:} the code model is shown the current implementation
and the failing checks and asked to return the corrected function.

\section{Representative natural bugs}\label{app:bugs}
This section makes the failure mode concrete by showing it in the models' own output.
Each listing below is a real one-shot, copied verbatim, next to the gold test case it
fails and the rule it breaks. We include them because the ``specification-completeness
omission'' we keep referring to is easier to trust when you can see the actual code:
these are ordinary bugs a strong code model writes when a ticket leaves its edge cases
unstated, not examples we invented. Every one has the same shape. The happy path is
right, and an invalid-input rule is skipped, which is exactly the region a free tester
never probes and an enumerated rule names directly.

\begin{lstlisting}[language=Python]
# split_money  -- fails (10, 0): returns [] instead of raising (R4)
def split(total_cents, n):
    if n <= 0:
        return []
    base = total_cents // n
    remainder = total_cents % n
    return [base + (1 if i < remainder else 0) for i in range(n)]

# paginate  -- fails ([1,2,3], 0, 2): no page / page_size validation (R2,R3)
def paginate(items, page, page_size):
    start = (page - 1) * page_size
    end = start + page_size
    return items[start:end]

# clamp  -- fails (5, 10, 0): no lo>hi check (R4)
def clamp(x, lo, hi):
    return max(lo, min(x, hi))

# parse_range  -- fails "5-1": no a>b check (R3)
def parse_range(s):
    if '-' in s:
        parts = s.split('-')
        start = int(parts[0]); end = int(parts[1])
        return list(range(start, end + 1))
    else:
        return [int(s)]
\end{lstlisting}

\noindent A further one-shot, \texttt{truncate}, shows the same omission, described here
rather than listed above. Rule R4 says any \texttt{max\_len} below 1 must raise, because
there is no room for even the ellipsis. The model skips that check and reuses the R3
formula \texttt{text[:max\_len-1]} plus the ellipsis; at \texttt{max\_len=0} the slice
\texttt{text[:-1]} is Python for ``all but the last character'', so it returns a trimmed
string with an ellipsis instead of raising. The strongest model in our study produced
omissions of the same kind.

\section{Pre-registered go/no-go}\label{app:prereg}
Before we looked at any results, we wrote down the exact bar the method had to clear to
count as a success, so we could not move the goalposts afterwards. The rule was that we
would declare \textbf{GO} only if every one of these held: the spec-over-free gain
$\Delta\ge+10$ pp; the 95\% bootstrap confidence interval excludes 0; $\Delta\ge+10$ on
$\ge2$ models; a detection advantage $\ge+15$ pp; no more than 5 pp of added false alarms;
and \spec{}@$K$ beats \free{}@$2K$ by $\ge+5$ pp. The observed results clear every
threshold.

\section{Additional results and controls}\label{app:extra}
This appendix collects the supporting checks referenced from the main results, in order: the
per-task detection breakdown, how many invalid-input tests each arm writes, a blind
self-refine control, a risk--coverage view that uses the test-pass fraction as a confidence
score, the cost accounting, a mutation-testing strength check, our AlphaCodium-style flow,
and what happens when the tester is shown the candidate code.

\paragraph{Per-task detection.} Table~\ref{tab:pertask} breaks core-task detection
down by task: \free{} catches none, \spec{} all, and \freeplus{} the obvious edges
(\texttt{paginate} 6/6) but not the non-obvious (\texttt{clamp} 2/6).

\begin{table}[tb]\centering
\caption{Detection on buggy one-shots, per task (core set; ``buggy'' out of 6).}
\label{tab:pertask}
\begin{tabular}{lcccc}
\toprule
Task & buggy & \free{} & \freeplus{} & \spec{}\\
\midrule
semver\_compare & 0 & -- & -- & --\\
median          & 0 & -- & -- & --\\
merge\_intervals& 0 & -- & -- & --\\
split\_money    & 6 & 0 & 3 & 6\\
paginate        & 6 & 0 & \textbf{6} & 6\\
truncate        & 6 & 0 & 4 & 6\\
roman\_to\_int  & 6 & 0 & 3 & 6\\
clamp           & 6 & 0 & \textbf{2} & 6\\
\midrule
total           & 30 & 0 & 18 & 30\\
\bottomrule
\end{tabular}
\end{table}

\paragraph{Negative space: invalid-input tests written per arm.} \free{} writes
\emph{zero} error-expecting tests on every task (so the bugs in the negative space go
untested); \freeplus{} writes some but fewer than \spec{}, especially where it
under-detects. Counts over six draws: \texttt{split\_money} 0/6/12,
\texttt{paginate} 0/8/12, \texttt{truncate} 0/4/6, \texttt{roman\_to\_int} 0/6/12,
\texttt{clamp} 0/2/6 (\free{}/\freeplus{}/\spec{}).

\paragraph{Blind self-refine (no tests).} Each code model re-reads its own one-shot
and the ticket and revises it, with no tests (the Self-Refine/Reflexion setting).
Final correctness moved from $38/38/54\%$ (plain) to $\mathbf{25/50/54}\%$ for the
small/medium/large model: on the small model it \emph{regressed three}
previously-correct functions while fixing none; on the medium model it fixed three; on
the large model it changed nothing net. Blind revision lands far below grounded tests
(100\%) and is not even monotone, consistent
with~\citet{huang2024selfcorrect} (self-correction can degrade) and
\citet{olausson2024selfrepair} (the missing ingredient is a good external feedback
signal, which the spec supplies).

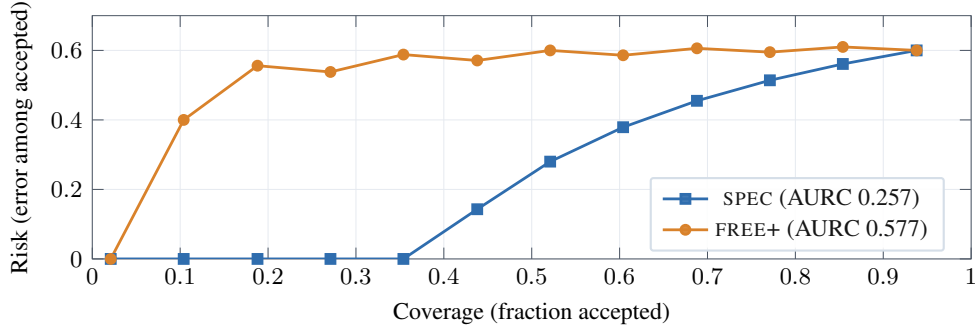
\begin{figure}[tb]\centering
\begin{tikzpicture}
\begin{axis}[width=0.8\linewidth, height=4.8cm, xlabel={Coverage (fraction accepted)},
  ylabel={Risk (error among accepted)}, xmin=0, xmax=1, ymin=0, ymax=0.7,
  xlabel style={font=\small}, ylabel style={font=\small},
  grid=both, major grid style={draw=plotgrid, line width=0.35pt},
  legend pos=south east,
  legend style={font=\footnotesize, fill=white, fill opacity=0.92, draw=codeframe, draw opacity=1},
  tick label style={font=\footnotesize}, thick, mark options={solid}]
\addplot[cspec, mark=square*] coordinates
  {(0.021,0.0)(0.104,0.0)(0.188,0.0)(0.271,0.0)(0.354,0.0)(0.438,0.143)(0.521,0.28)(0.604,0.379)(0.688,0.455)(0.771,0.514)(0.854,0.561)(0.938,0.6)};
\addplot[cfreeplus, mark=*] coordinates
  {(0.021,0.0)(0.104,0.4)(0.188,0.556)(0.271,0.538)(0.354,0.588)(0.438,0.571)(0.521,0.6)(0.604,0.586)(0.688,0.606)(0.771,0.595)(0.854,0.61)(0.938,0.6)};
\legend{\spec{} (AURC 0.257), \freeplus{} (AURC 0.577)}
\end{axis}
\end{tikzpicture}
\caption{Risk--coverage curves using each arm's test-pass fraction as a confidence
score. Lower is better.}
\label{fig:rc}
\end{figure}

\paragraph{Risk--coverage (selective prediction).} Figure~\ref{fig:rc} uses each arm's
test-pass fraction as a confidence score for ``accept this code.'' \spec{}'s curve
hugs zero risk out to $\sim\!35\%$ coverage; \free{}'s risk is high immediately
(AURC 0.257 vs.\ 0.577).

\paragraph{Cost.} Estimated per-task cost ($\approx\!4$ chars/token; exact call
counts; list prices): Sonnet plain 1 call, \$1.84/1k tasks, 38\%; Sonnet$+$\freeplus{}
3 calls, \$5.83/1k, 62\%; Sonnet$+$framework 3 calls, \$6.84/1k, 100\%; Opus plain 1
call, \$3.07/1k, 54\%. Grounding costs about the same as the fair baseline yet adds
$38$\pp{}; capability substitution costs $\approx\!2.2\times$ a plain Opus call.

\paragraph{Mutation testing (suite strength).} As a structural strength check, we
mutate each trusted reference (relational-op boundary flips, arithmetic-op swaps,
integer-constant $\pm1$, and deletion of \texttt{if\,\dots:\,raise} guards), keep the
non-equivalent mutants (those that differ from the reference on the gold suite), and
measure what fraction each arm's generated suite kills. Over 10 tasks (6 scale-up
$+$ 4 held-out; 147 non-equivalent mutants), mutation score is \spec{} 93.9\%
(138/147) and \freeplus{} 95.9\% (141/147), and it does \emph{not} favour \spec{}. The
reason is instructive: mutation score rewards a suite for being \emph{sensitive}
regardless of whether its expectations are \emph{correct}. On \texttt{format\_duration}
the \freeplus{} suite kills 18/18 vs.\ \spec{}'s 14/18 precisely \emph{because} its
out-of-spec assertions (``1h 0m 0s'') fail on more mutants, yet those same assertions
reject the correct reference and misguide repair (\freeplus{} final correctness 0\%
there, \S\ref{sec:scaleup}). A suite can be mutation-strong yet wrong; this is exactly
why our main metric is end-to-end final correctness, not mutation score.

\paragraph{AlphaCodium-style flow: what we implemented.} Our \S\ref{sec:prop} flow
re-creates the core loop of~\citet{ridnik2024alphacodium} on single-function tasks:
(1) the model reflects on the ticket and lists requirements, including edges and
invalid inputs; (2) it generates its own AI tests from that reflection (a $2K$
budget, larger than \spec{}); (3) it writes a solution; (4) it iterates up to three
rounds, at each step re-running its AI tests and \emph{dual-fixing} both the code and
the tests (it may revise a test it judges wrong). It is given only the ticket, never
the enumerated rules. This captures the published flow's defining ingredients
(problem reflection, self-generated tests, iterative code/test repair) but is
\emph{not} the original system: we omit public-test reasoning and multi-candidate
solution ranking, and we did not re-tune it against the original's benchmark numbers.
So this is a faithful re-creation of the flow's structure rather than a validated
port; the full runnable implementation (\texttt{alphacodium.py}) is complete and
self-contained, so the comparison can be audited or strengthened. The takeaway is robust to these caveats:
even with a larger test budget and an iterative dual-fix loop, a flow that lacks the
enumerated spec lands at the \freeplus{} level (72\%), not \spec{}'s 100\%.

\paragraph{When the tester sees the code.} Our tester writes tests from the
ticket/rules, never from the candidate code, because a tester shown buggy code tends to
certify the bug as intended. We verified this directly: re-generating tests for 24 buggy
one-shots with the candidate code \emph{included} in the prompt, grounded detection is
\emph{unchanged} (\spec{} 24/24 $\to$ 24/24) while the ungrounded baseline
\emph{collapses to zero} (\freeplus{} 7/24 $\to$ 0/24). Shown the code, the free tester writes tests that match
the buggy behaviour; the spec's rules still contradict it. This is a third face of the
same mechanism: grounding anchors the expected values to the specification, whereas an
ungrounded tester anchors them to whatever code it is shown.

\section{Per-model detail}\label{app:permodel}
Table~\ref{tab:permodel} gives the full per-model breakdown on the eight core
tasks (3 seeds; detection on buggy one-shots; \freeplus{} with two repair rounds).
The grounded arm catches every buggy one-shot on every model; \freeplus{} detection
\emph{falls} as the code model gets stronger (its bugs get subtler), yet \spec{}
still reaches 100\% everywhere.

\begin{table}[tb]\centering
\caption{Per-model results on the 8 core tasks. ``det'' = detection on buggy
one-shots (caught/buggy); ``fin'' = final correctness.}
\label{tab:permodel}
\begin{tabular}{lcccccc}
\toprule
Model & plain & self-refine & \freeplus{} det & \freeplus{} fin & \spec{} det & \spec{} fin\\
\midrule
Claude Haiku 4.5  & 38\% & 25\% & 11/15 & 62\% & 15/15 & 100\%\\
Claude Sonnet 4.6 & 38\% & 50\% & 7/15  & 62\% & 15/15 & 100\%\\
Claude Opus 4.8   & 54\% & 54\% & 2/11  & 62\% & 11/11 & 100\%\\
\bottomrule
\end{tabular}
\end{table}

\section{Cross-vendor replication}\label{app:xvendor}
A fair worry is that the effect rides on something specific to Anthropic's models, since
the main experiments use Claude for most roles. This section reruns the whole pipeline
with non-Anthropic families to rule that out. Each model plays every role \emph{full-stack}
(coder, tester, and repairer), so if the grounding gap shows up here it cannot depend on
any Anthropic component. Under the hood, the model client routes each request by model id
to Anthropic, OpenAI, or Google, and reads back a parsed JSON object and normalised token
counts through each provider's own structured-output mechanism, at each model's default
settings.

\textbf{Results} (12 specification-completeness tasks, 3 seeds, 36 instances per
model). Both non-Anthropic families reproduce the grounding gap (Table~\ref{tab:xvendor});
detection on buggy one-shots is \free{} 5/18 (codex) / 7/24 (Gemini), \freeplus{}
16/18 / 20/24, \spec{} 18/18 / 24/24, the same detection-then-repair structure seen
on Claude, where \freeplus{} detects nearly all bugs but only some of its repairs
reach the gold oracle while \spec{} detects and repairs (almost) all.

\begin{table}[tb]\centering
\caption{Cross-vendor replication, full-stack (each model is coder $=$ tester $=$
repairer). The grounding gap reproduces on both non-Anthropic families. Claude tiers
are Haiku~4.5 / Sonnet~4.6 / Opus~4.8; the non-Anthropic models are GPT-5.3-codex
(OpenAI) and Gemini~3.5~Flash (Google).}
\label{tab:xvendor}
\setlength{\tabcolsep}{5pt}
\begin{tabular}{lcccc}
\toprule
Family & natural bug rate & \freeplus{} final & \spec{} final & \spec{}$-$\freeplus{}\\
\midrule
Claude (3 tiers)          & 62\%/100\%\,(core/held) & 62\% & 100\% & $+38$\\
GPT-5.3-codex (OpenAI)    & 50\% & 72\% & 100\% & $+28$\\
Gemini 3.5 Flash (Google) & 67\% & 72\% & 92\%  & $+19$\\
\bottomrule
\end{tabular}
\end{table}

\textbf{Generation settings and cross-vendor parity.} We run code generation as a
single quick ``fast pass'' (the regime that elicits natural specification-completeness
bugs), \emph{without} extended reasoning, consistently across families. This matters
because Gemini~3.5~Flash enables a ``thinking'' mode by default that spends far more
output-billed reasoning tokens per call than GPT-5.3-codex ($\approx\!260$ output
tokens/call, including its light Responses-API reasoning) or the Claude code models
(left at their default reasoning effort). Left on, Gemini is the \emph{outlier} in
reasoning effort, and because thinking tokens are billed as output it also truncated
the JSON response and inflated cost; disabling it brings all three families into the
same light-reasoning regime. The reasoning setting is held \emph{constant within each vendor}
across \spec{} and \freeplus{}, so it cannot bias the grounding \emph{gap} we
report; it affects only each family's absolute level, which we do not compare across
families. We disclose it for transparency.

\section{External validity: porting to HumanEval+ (EvalPlus)}\label{app:evalplus}
To test whether the grounding effect is an artifact of our bespoke suite, we port the
\emph{entire} pipeline to a public benchmark judged by a public, independently-authored
oracle. Tasks are 24 HumanEval problems sampled evenly across HumanEval/0--163 (a
systematic stride, not the easy early ones; restricted to the exact-comparable problems).
Both final correctness and detection are judged by \emph{EvalPlus's own} expanded test
suite \citep{liu2023evalplus}: the released canonical solution evaluated over EvalPlus
base${+}$plus inputs ($\approx 203$ cases per task). Neither the benchmark nor the oracle
is ours; the only authored artifact is the \spec{} rule enumeration, frozen once from each
public docstring (mean $K=6.1$). We run full-stack (coder ${=}$ tester ${=}$ repairer) on
five models: GPT-5.3-codex and Gemini~3.5~Flash via API, and the three Claude tiers via
single-pass app inference with no code execution.

\begin{table}[tb]
\centering
\caption{HumanEval+ port, full-stack, judged by EvalPlus's own oracle. \spec{} never
outdetects \freeplus{}: the grounding gap is absent because the benchmark elicits logic,
not specification-completeness, defects.}
\label{tab:evalplus}
\setlength{\tabcolsep}{6pt}
\begin{tabular}{lccc}
\toprule
Model (full-stack) & one-shot bug rate & \spec{} detect & \freeplus{} detect\\
\midrule
GPT-5.3-codex (OpenAI)        & 8\%\,(6/72)  & 5/6 & 6/6\\
Gemini 3.5 Flash (Google)     & 3\%\,(2/72)  & 0/2 & 0/2\\
Claude Haiku 4.5 (Anthropic)  & 12\%\,(9/72) & 8/9 & 8/9\\
Claude Sonnet 4.6 (Anthropic) & 4\%\,(3/72)  & 0/3 & 0/3\\
Claude Opus 4.8 (Anthropic)   & 3\%\,(2/72)  & 0/2 & 0/2\\
\midrule
\textbf{Pooled}               & \textbf{6\%\,(22/360)} & \textbf{13/22} & \textbf{14/22}\\
\bottomrule
\end{tabular}
\end{table}

The grounding gap vanishes, as our scope predicts. Frontier models rarely produce
specification-completeness defects on HumanEval+ (one-shot bug rate 3--12\%): the problems
are well-specified, with worked examples and valid-input assumptions, so the bugs that do
occur are ordinary \emph{logic} errors (e.g.\ HumanEval/107's \texttt{range(1,n)}
off-by-one), which any exercising test catches without the spec's enumerated edge rules.
Pooled over 360 one-shots, \spec{} and \freeplus{} detect the 22 buggy ones comparably
(13 vs.\ 14, \freeplus{} marginally ahead), and on no individual model does \spec{}
outdetect \freeplus{}; across the Claude tiers their false-alarm rates are identical
(38/202 each). This is the logic-task null of \S\ref{sec:capability} reproduced on
a public benchmark with a public oracle. It \emph{bounds} the claim rather than extending
it: the specification-completeness bug class the method targets is largely absent from
well-specified algorithmic benchmarks, which is exactly why eliciting and measuring the
effect calls for realistic, under-specified tickets. The bespoke suite is necessary to
exhibit the failure mode, not favorable to the method.

\section{Additional robustness on an independent inference stack}\label{app:copilot}
The experiments in the main paper use direct vendor APIs (Anthropic, OpenAI,
Google). As an independent check, we ran nine further experiments on a different
inference stack, the VS Code Copilot model API, which serves Anthropic and OpenAI
models through one interface. They let us vary the \emph{tester} model widely
(\S\ref{app:e2}), ask whether the specification must be hand-written (\S\ref{app:e1}),
measure how much specification is enough (\S\ref{app:e3}), reproduce the grounding
benefit on new tasks with code from a different stack (\S\ref{app:b}), test tasks
whose edges are fixed by an external standard (\S\ref{app:rw}), scale the suite up
while adding stateful, multi-step tasks (\S\ref{app:scale2}), reimplement
standard-library functions judged against real CPython (\S\ref{app:std}), reimplement
functions vendored verbatim from real ML repositories (transformers, NLTK; \S\ref{app:repo}),
and reimplement the PyPA \texttt{packaging} library's version utilities judged against the
real installed library (\S\ref{app:pkg}).
The first three reuse the paper's frozen one-shots and judge every test by the same trusted
reference and gold oracle, so the numbers are comparable; the last six add fresh tasks and
one-shots. One
caveat: this API does not enforce structured output, so we add a tolerant parser that
also accepts single-quoted JSON. The fixed strong tester (Claude Sonnet 4.6) used in the
spec-quality experiments below is unaffected by parsing; only the weakest models in the
tester sweep lean on the tolerant parser.

\subsection{The grounding benefit holds across tester capability}\label{app:e2}
\begin{figure}[tb]\centering
\begin{tikzpicture}
\begin{axis}[width=0.47\textwidth, height=4.6cm, title={\footnotesize detection (\%)},
  ymin=42, ymax=107, ytick={50,75,100},
  xtick={1,2,3,4,5,6}, xticklabels={4o-mini,5-mini,codex,5.5,Sonnet,Opus},
  x tick label style={rotate=35, anchor=east, font=\scriptsize},
  legend pos=south west,
  legend style={font=\scriptsize, fill=white, fill opacity=0.92, draw=codeframe, draw opacity=1},
  grid=both, major grid style={draw=plotgrid, line width=0.35pt}, thick, mark options={solid}]
\addplot[cfreeplus, mark=*] coordinates {(1,73)(2,80)(3,77)(4,93)(5,100)(6,50)};
\addplot[cspec, mark=square*] coordinates {(1,93)(2,100)(3,100)(4,100)(5,100)(6,100)};
\legend{\freeplus{}, \spec{}}
\end{axis}
\end{tikzpicture}\hfill
\begin{tikzpicture}
\begin{axis}[width=0.47\textwidth, height=4.6cm, title={\footnotesize false alarm (\%)},
  ymin=-6, ymax=98, ytick={0,25,50,75},
  xtick={1,2,3,4,5,6}, xticklabels={4o-mini,5-mini,codex,5.5,Sonnet,Opus},
  x tick label style={rotate=35, anchor=east, font=\scriptsize},
  legend pos=north east,
  legend style={font=\scriptsize, fill=white, fill opacity=0.92, draw=codeframe, draw opacity=1},
  grid=both, major grid style={draw=plotgrid, line width=0.35pt}, thick, mark options={solid}]
\addplot[cfreeplus, mark=*] coordinates {(1,89)(2,56)(3,67)(4,56)(5,11)(6,0)};
\addplot[cspec, mark=square*] coordinates {(1,67)(2,44)(3,0)(4,0)(5,11)(6,0)};
\legend{\freeplus{}, \spec{}}
\end{axis}
\end{tikzpicture}
\caption{Tester sweep (weak $\to$ strong), \freeplus{} vs.\ \spec{}. Left: detection on
the 30 buggy one-shots; \spec{} (blue) is at or above \freeplus{} (orange) on every
tester. Right: false alarms on the 18 correct one-shots; \spec{} is at or below
\freeplus{} everywhere, with the largest precision gains in the middle of the range
(codex, 5.5). The two smallest models keep some \spec{} false alarms (the capability
floor); the largest, Opus, gains the most detection.}
\label{fig:e2}
\end{figure}
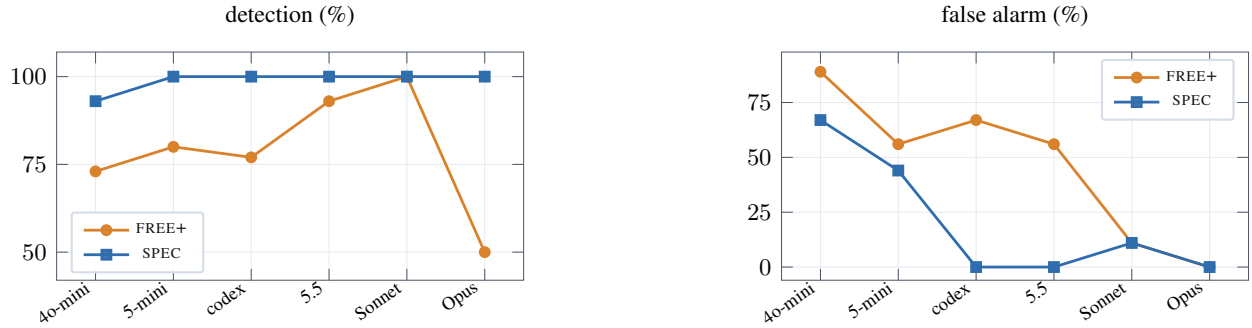

The main paper varies the \emph{coder} tier (\S\ref{sec:capability}) but holds the
tester fixed at Sonnet~4.6. Here we vary the \emph{tester} instead: we hold the
frozen one-shots fixed and re-author tests with six models spanning two vendors and
five capability tiers, then score detection on the 30 buggy core one-shots and
false alarms on the 18 correct ones (Table~\ref{tab:e2}, Figure~\ref{fig:e2}). Spec detection is at least
as high as \freeplus{} on every one of the six testers (gaps from $+0$ to $+50$\pp),
and spec false alarms are no higher than \freeplus{} on every tester. The precision
win is largest in the middle of the range: GPT-5.3-codex and GPT-5.5 drop from
$56$--$67\%$ false alarms under \freeplus{} to $0\%$ under \spec{}. There is an
honest floor: the two smallest models (GPT-4o-mini, GPT-5-mini) still emit some
wrong expected values even when given the rules, so their spec false alarms stay at
$44$--$67\%$. Grounding helps once the tester is capable enough to apply a rule
correctly, and never hurts.

\begin{table}[tb]\centering
\caption{Tester sweep on frozen one-shots (Copilot API). Detection on 30 buggy
one-shots; false alarms on 18 correct one-shots; tester accuracy is the share of
drawn cases whose expected value matches the trusted reference. \spec{} detection
$\ge$ \freeplus{} and \spec{} false alarms $\le$ \freeplus{} on every tester.}
\label{tab:e2}
\begin{tabular}{lcccccc}
\toprule
& \multicolumn{2}{c}{detection} & \multicolumn{2}{c}{false alarm} & \multicolumn{2}{c}{tester acc}\\
\cmidrule(lr){2-3}\cmidrule(lr){4-5}\cmidrule(lr){6-7}
Tester (weak $\to$ strong) & \freeplus{} & \spec{} & \freeplus{} & \spec{} & \freeplus{} & \spec{}\\
\midrule
GPT-4o-mini        & 73\%  & 93\%  & 89\% & 67\% & 72\%  & 77\%\\
GPT-5-mini         & 80\%  & 100\% & 56\% & 44\% & 88\%  & 88\%\\
GPT-5.3-codex      & 77\%  & 100\% & 67\% & 0\%  & 86\%  & 99\%\\
GPT-5.5            & 93\%  & 100\% & 56\% & 0\%  & 90\%  & 100\%\\
Claude Sonnet 4.6  & 100\% & 100\% & 11\% & 11\% & 97\%  & 98\%\\
Claude Opus 4.8    & 50\%  & 100\% & 0\%  & 0\%  & 100\% & 100\%\\
\bottomrule
\end{tabular}
\end{table}

\subsection{What automatically derived specifications miss}\label{app:e1}
A fair objection is that the rules are hand-written, so the method just moves the
work. We test whether a model can derive them. For each core task we give Sonnet~4.6
\emph{only the ticket and signature} (never the reference or the gold suite) and ask
it to enumerate the same number of rules as the human spec. We freeze those
auto-derived rules, then run tester arms on the frozen one-shots, all with the
same fixed tester (Sonnet~4.6) and all judged by the gold oracle: \freeplus{} (no
rules), \spec{}@human (the hand-written rules), \spec{}@auto (rules a plain prompt
derives), and \spec{}@auto+ (rules a prompt that explicitly asks for error and boundary
conditions derives).

The derived rules \emph{read} well, but they recover only part of the detection
(Table~\ref{tab:e1}): \spec{}@auto catches $47\%$ of the bugs that \spec{}@human
catches, and the loss is not spread evenly. Per task, \spec{}@auto matches the human
spec on \texttt{paginate} (18/18) and nearly on \texttt{split\_money} (15/18), but
falls to \textbf{0/18} on both \texttt{roman\_to\_int} and \texttt{clamp}. Its tests
are still accurate (0 false alarms, $96\%$ tester accuracy); they simply do not
\emph{trigger} the bug. The reason is precise, and it echoes the paper's central point:
the auto-generator paraphrases the behaviour the ticket \emph{states} (the
happy path) but does not invent the conditions the ticket leaves \emph{unstated},
which is exactly the specification-completeness bug class. For \texttt{clamp}, the
human spec's rule ``if \texttt{lo > hi}, raise \texttt{ValueError}'' names the bug; the
auto spec instead spent its fourth rule on the benign ``if \texttt{lo == hi}, return
\texttt{lo}'' and never probed \texttt{lo > hi}. For \texttt{roman\_to\_int}, all five
auto rules describe parsing of valid input, with no rule for the empty string or an
invalid character, so the validation bug goes untested. The specification's value is
concentrated in the curated edge and error rules the ticket omits; a model can restate
a ticket, but restating it does not surface what the ticket forgot.

So we test the obvious fix directly. \spec{}@auto+ uses the same generator and the same
ticket, but the prompt now asks it to prioritise the error and boundary conditions
(\S\ref{app:e3} shows these carry the detection signal). Detection fully recovers:
\spec{}@auto+ reaches $100\%$, matching the human spec and catching every bug on
\texttt{roman\_to\_int} and \texttt{clamp} that the plain version missed. But the recovery
comes at a price. Forced to enumerate edges, the generator now \emph{invents} error
semantics the ticket never pins down, and some are wrong: its \texttt{roman\_to\_int}
rules declare the parser case-sensitive, so a test rejects \texttt{roman\_to\_int("iv")},
which the reference accepts. False alarms jump from $0\%$ to $33\%$ and tester accuracy
falls to $93\%$. So the detection half of the specification automates with the right
prompt, but the precision half does not: getting the edges \emph{right}, not merely
listing them, is the curation a human spec still supplies. Pushed to cover the edges
blind, the auto-generator re-enters the same invent-the-oracle failure mode as the
ungrounded tester (\S\ref{sec:falsealarm}). (With this strong tester \freeplus{}
detection also ceilings at $100\%$, so the informative contrast is among the spec
variants, not the \freeplus{} baseline.)

\begin{table}[tb]\centering
\caption{Auto-derived vs.\ hand-written specifications, same fixed tester (Sonnet 4.6)
on the frozen one-shots, judged by the gold oracle. A plain prompt (\spec{}@auto) stays
accurate but recovers only $47\%$ of \spec{}@human detection, missing the \emph{unstated}
error conditions on \texttt{roman\_to\_int} and \texttt{clamp}. Prompting for the edges
(\spec{}@auto+) recovers all detection but invents some wrong error semantics, so false
alarms rise to $33\%$: detection automates, precision still needs human curation.}
\label{tab:e1}
\begin{tabular}{lccc}
\toprule
Tester arm & detection & false alarm & tester acc\\
\midrule
\freeplus{} (no rules)          & 100\% & 0\%  & 98\%\\
\spec{}@human (hand-written)    & 100\% & 0\%  & 99\%\\
\spec{}@auto (plain prompt)     & 47\%  & 0\%  & 96\%\\
\spec{}@auto+ (edge-prioritised)& 100\% & 33\% & 93\%\\
\bottomrule
\end{tabular}
\end{table}

\paragraph{Spec authoring cost.} The specs are short: $4.75$ rules per core task on
average ($38$ rules over $8$ tasks), and an LLM drafts most of the text for free (the
\spec{}@auto column). What an auto-draft cannot supply is the small set of \emph{error
and boundary} rules it omits or invents: on the eight core tasks a human need add or
correct just five such rules (one on \texttt{clamp}, two on \texttt{roman\_to\_int}, one
on \texttt{truncate}, one on \texttt{split\_money}), and those are exactly the four tasks
where \spec{}@auto detection collapses (Table~\ref{tab:e1}). So the human effort is small
in volume but high in leverage: curate a handful of failure-mode rules, not write the
whole spec. Those rules come from sources a developer already has: the invalid-input
policy, type and domain preconditions, boundary conventions, and any external contract (an
RFC, a format spec, a library's documented errors), which is exactly where the real-world
slice (\S\ref{app:rw}) takes its edges.

\subsection{Detection rises with specification coverage}\label{app:e3}
\begin{figure}[tb]\centering
\begin{tikzpicture}
\begin{axis}[
  width=0.64\textwidth, height=4.8cm,
  xlabel={specification coverage given to the tester}, ylabel={detection (\%)},
  xtick={1,2,3}, xticklabels={min (1 rule), half, full},
  xmin=0.85, xmax=3.15, ymin=-6, ymax=108, ytick={0,25,50,75,100},
  legend pos=south east, legend cell align=left,
  legend style={font=\footnotesize, fill=white, fill opacity=0.92, draw=codeframe, draw opacity=1},
  grid=both, major grid style={draw=plotgrid, line width=0.35pt}, thick, mark options={solid}]
\addplot[cfreeplus, mark=*, line width=1pt] coordinates {(1,0) (2,30) (3,100)};
\addplot[cspec, mark=square*, line width=1pt] coordinates {(1,70) (2,100) (3,100)};
\legend{forward (validation rules last), edges-first (validation rules first)}
\end{axis}
\end{tikzpicture}
\caption{Specification dose-response: detection on the 30 buggy core one-shots as the
spec tester is given a growing subset of rules. Forward order rises
$0/60\to18/60\to60/60$ as coverage grows; edges-first already reaches $42/60$ ($70\%$)
with a single validation rule and $60/60$ by half coverage. The \emph{content} of the
validation rules, not the rule count, drives detection.}
\label{fig:e3}
\end{figure}
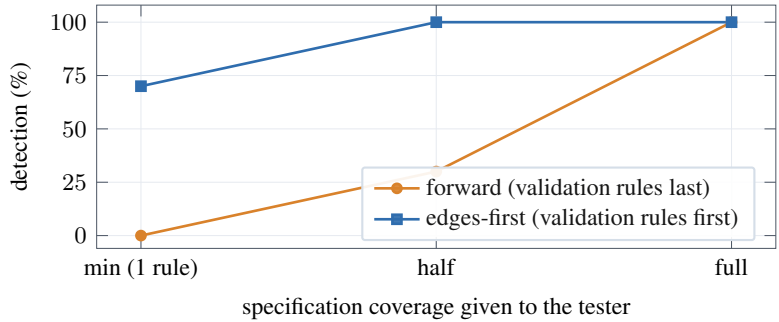

Finally we measure dose-response: how detection scales with how much of the
specification the tester is given. For each core task we run the spec tester with a
growing subset of its rules, under two orderings: \emph{forward} (rules in their
natural order, where validation rules usually come last) and \emph{edges-first}
(validation and error rules moved to the front). The pattern is clean
(Figure~\ref{fig:e3}). Under forward order, detection rises monotonically with
coverage, $0\%\to30\%\to100\%$: one happy-path rule catches nothing, and you need the
full specification to catch every bug. Under edges-first order, detection is already
$70\%$ with a \emph{single} rule and $100\%$ at half coverage. So detection scales
with coverage of the \emph{right} rules, not with the raw rule count: the validation
rules that name the failure mode carry the signal, which is why a complete
specification, or at least one that front-loads its error conditions, drives
detection.

\subsection{Replication on new tasks with a fresh code stack}\label{app:b}
The experiments above reuse the paper's frozen one-shots, so a fair worry is that the
effect is tied to our task suite or to how that original code was written. We therefore
authored five \emph{new} specification-completeness functions after the fact
(\texttt{to\_ordinal}, \texttt{time\_to\_minutes}, \texttt{nth\_triangular},
\texttt{parse\_hex\_byte}, \texttt{roman\_from\_int}), each with an anchor-validated
reference oracle, and generated fresh one-shots for them with three Copilot coder models
(GPT-5-mini, GPT-5.3-codex, Sonnet~4.6; two seeds each, 30 one-shots). The new tickets
elicit the same defect class: the one-shot bug rate is $67\%$ (20/30), with coders
omitting the unstated error and boundary behaviour (for instance, every
\texttt{nth\_triangular} one-shot forgot to reject a negative index). On the buggy
one-shots a fixed Sonnet~4.6 tester detects $95\%$ (57/60) under \spec{} versus $57\%$
(34/60) under \freeplus{}, a $+38$\pp{} gap, with zero false alarms on either arm
(Table~\ref{tab:b}). The effect reproduces at full strength on tasks we wrote after the
study and on code from a different inference stack, not only on the original frozen
suite.

\begin{table}[tb]\centering
\caption{Five new validation tasks, fresh one-shots from three Copilot coders (30
one-shots, $67\%$ buggy), fixed Sonnet 4.6 tester, judged by anchor-validated gold
oracles. \spec{} detects $+38$\pp{} more of the buggy one-shots than \freeplus{}, with no
false alarms on either arm.}
\label{tab:b}
\begin{tabular}{lccc}
\toprule
Tester arm & detection & false alarm & tester acc\\
\midrule
\freeplus{} & 57\% (34/60) & 0\% (0/30) & 98\%\\
\spec{}     & 95\% (57/60) & 0\% (0/30) & 100\%\\
\bottomrule
\end{tabular}
\end{table}

\subsection{A real-world slice: tasks fixed by external contracts}\label{app:rw}
A sharper test of external validity uses tasks whose correct edge behaviour is set by a
\emph{published standard or a widely used library}, not by us, so the oracle cannot be
``favourable by design.'' We authored six such tasks (strict dotted-quad IPv4 parsing per
RFC~791, CSS hex-colour parsing, ISO ISBN-10 checksum validation, CamelCase-to-snake
conversion, compact duration parsing, and URL slugification), each with an anchor-validated
reference that matches the external contract and an under-specified ticket that states the
happy path but omits the edges. Fresh one-shots from three Copilot coders (two seeds; 34
valid one-shots) reproduce the defect class at full force: the one-shot bug rate is $85\%$
(29/34), with coders omitting the documented edges (every \texttt{parse\_ipv4},
\texttt{validate\_isbn10}, and \texttt{parse\_duration} one-shot is buggy). A fixed
Sonnet~4.6 tester then detects $\mathbf{100\%}$ (87/87) of the bugs under \spec{} versus
$\mathbf{32\%}$ (28/87) under \freeplus{}, a $+68$\pp{} gap (the largest in the paper),
with zero false alarms on either arm and tester accuracy rising $91\%$ to $100\%$
(Table~\ref{tab:rw}). The gap is \emph{larger} here than on the core suite precisely
because real contracts hide more unstated edges (leading-zero IPv4 octets, an acronym
run in CamelCase such as \texttt{HTTPServer}, a bare number with no duration unit) that an
edge-prompted but ungrounded tester does not think to probe. The effect is not an artifact of tasks we
designed to show it.

\begin{table}[tb]\centering
\caption{Real-world slice: six tasks whose edge behaviour is fixed by external standards
(RFC, CSS, ISO ISBN-10, common library conventions); fresh one-shots from three Copilot
coders (34 one-shots, $85\%$ buggy); fixed Sonnet 4.6 tester, judged by anchor-validated
oracles. The $+68$\pp{} detection gap is the largest in the paper.}
\label{tab:rw}
\begin{tabular}{lccc}
\toprule
Tester arm & detection & false alarm & tester acc\\
\midrule
\freeplus{} & 32\% (28/87) & 0\% (0/15) & 91\%\\
\spec{}     & 100\% (87/87) & 0\% (0/15) & 100\%\\
\bottomrule
\end{tabular}
\end{table}

\paragraph{Best-of-$N$: more sampling does not substitute for grounding.} A natural rescue
for the ungrounded arm is to draw \emph{many} \freeplus{} suites and union them. Drawing
eight independent \freeplus{} suites per task and unioning the first $k$, detection on the
29 buggy one-shots saturates almost at once: $9/29$ at $k{=}1$, then $15/29$ ($52\%$) from
$k{=}2$ onward, \emph{flat} all the way to $k{=}8$ (Table~\ref{tab:bon}). The 14 bugs
\freeplus{} never catches are the most contract-specific edges (leading-zero IPv4 octets,
acronym runs in CamelCase-to-snake conversion, a bare number with no duration unit): an
ungrounded tester re-probes the same obvious cases and never invents the missing rule, so
extra samples add nothing, while \spec{} reaches $100\%$ with a single grounded draw.

\begin{table}[tb]\centering
\caption{Best-of-$N$ on the real-world slice (29 buggy one-shots): detection when the first
$k$ independent \freeplus{} suites are unioned (caught out of 29). Ungrounded sampling
plateaus at $52\%$ from the second draw; a single grounded \spec{} draw reaches $100\%$.}
\label{tab:bon}
\begin{tabular}{lccccc}
\toprule
& \freeplus{} $k{=}1$ & \freeplus{} $k{=}2$ & \freeplus{} $k{=}4$ & \freeplus{} $k{=}8$ & \spec{}\\
\midrule
Detection & 31\% (9) & 52\% (15) & 52\% (15) & 52\% (15) & \textbf{100\% (29)}\\
\bottomrule
\end{tabular}
\end{table}

\subsection{Scaling up and diversifying the task suite}\label{app:scale2}
Two external-validity questions remain: whether the effect is tied to the \emph{size} of
the suite, and whether it depends on its tasks being \emph{single pure functions}. We add
sixteen fresh tasks,
authored after the study and anchor-validated, and run the full pipeline on the
independent Copilot stack (two Copilot coders for the one-shots; fixed Sonnet 4.6
tester), growing the treatment base from 18 to 34. Ten are more single-function
specification-completeness utilities (\texttt{iso\_weekday}, \texttt{parse\_int\_base},
\texttt{validate\_isbn13}, \texttt{snake\_to\_camel}, and so on); six are
\emph{stateful or multi-step}: the entry point consumes an \emph{operation sequence}
and correctness depends on accumulated state and cross-operation contracts (a bank ledger
that must reject overdrafts, a bounded stack that must reject overflow and underflow, a
token-bucket limiter, an id pool, a phone normaliser, an HTTP query parser), a different
bug surface from single-call validation. The natural one-shot bug rate is high on both
($60\%$ single-function, $96\%$ stateful, $73\%$ overall), so the defect class transfers
to the new tasks.

The grounding effect reproduces on both slices (Table~\ref{tab:scale2}). On the
stateful/multi-step slice, detection rises from \freeplus{} $83\%$ to \spec{} $100\%$
($+17$\pp{}): grounding closes the cross-operation contracts (overdraft, overflow,
exhaustion) that an edge-prompted but ungrounded tester only probes in part. On the ten
new single-function tasks detection is near the ceiling for both (\freeplus{} $96\%$,
\spec{} $100\%$), and the benefit lands where the scale-up section predicted
(\S\ref{sec:scaleup}), in \emph{precision}: \freeplus{} falsely rejects correct code on
$6\%$ of checks at $94\%$ tester accuracy, while \spec{} holds $0\%$ and $100\%$. One
honest wrinkle: on the stateful tasks the grounded tester's own accuracy dips to $87\%$,
because the exact output \emph{sequence} for a multi-operation input is hard to work out
by hand; this does not turn into a false alarm (the single correct stateful one-shot is
never wrongly rejected) and detection still reaches $100\%$, because the error-contract
cases that carry detection (\S\ref{app:e3}) are the easy ones to get right. So the effect
holds on a larger suite and on stateful, multi-step tasks, not only single pure
functions.

\begin{table}[tb]\centering
\caption{Scaling and diversifying the suite (independent Copilot stack; two Copilot
coders; fixed Sonnet 4.6 tester, 2 draws; gold-oracle judged). DIV is six stateful or
multi-step tasks (the entry point consumes an operation sequence); N is ten more
single-function tasks. Detection is on buggy one-shots, false alarms on correct ones.}
\label{tab:scale2}
\begin{tabular}{llccc}
\toprule
Slice & arm & detection & false alarm & tester acc\\
\midrule
DIV: stateful/multi-step & \freeplus{} & 83\% (38/46)  & 0\% (0/2)  & 97\%\\
\;(23 buggy one-shots)   & \spec{}     & 100\% (46/46) & 0\% (0/2)  & 87\%\\
\midrule
N: more single-function  & \freeplus{} & 96\% (46/48)  & 6\% (2/32) & 94\%\\
\;(24 buggy one-shots)   & \spec{}     & 100\% (48/48) & 0\% (0/32) & 100\%\\
\bottomrule
\end{tabular}
\end{table}

\subsection{A real external oracle: reimplementing the standard library}\label{app:std}
The sharpest answer to ``the oracle is favourable by design'' is an oracle we did not
write. We take eight Python standard-library utilities (whitespace-collapsing truncation,
word titlecasing, shell-style argument splitting, sign-aware zero-padding, URL
percent-encoding, the last path segment, tab expansion, and path normalisation) and ask a
coder to reimplement each from an \emph{under-specified} ticket under a \emph{neutral}
name, so it cannot simply recall the library. Correctness, detection, and false alarms are
all judged against the \emph{actual CPython function} (\texttt{textwrap.shorten},
\texttt{shlex.split}, \texttt{str.zfill}, and so on): the oracle is battle-tested
production code, and the \spec{} rules restate the library's own documented behaviour, not
a contract of ours. The bug class survives this real oracle, with a $47\%$ one-shot bug
rate (a first version that \emph{named} the library function was reproduced almost
perfectly, the EvalPlus null again, which is why the tickets are under-specified).

Three results follow (Table~\ref{tab:std}). \emph{Grounding} (\#1): against the real
oracle, \spec{} detects $100\%$ of the bugs at $0\%$ false alarms, while the edge-prompted
\freeplus{} detects $83\%$ but falsely rejects correct code $68\%$ of the time. Not knowing
the exact documented behaviour (how \texttt{textwrap.shorten} collapses whitespace, how
\texttt{shlex} treats a backslash), the ungrounded tester invents wrong expected values and
rejects implementations that in fact match CPython; this is the largest false-alarm gap in
the paper outside the saturated \texttt{packaging} slice (App.~\ref{app:pkg}), on an oracle
we did not author. \emph{Coverage is not grounding} (\#3): a third
arm prompted to \emph{maximise branch and path coverage} catches only $37\%$, \emph{below}
even \freeplus{} -- it writes accurate but shallow cases (94\% accuracy, $0\%$ false
alarms) that execute paths without asserting the contract value on the error conditions, so
it misses the specification bugs. \emph{A weak tester} (\#2): repeating with a cheap
GPT-4o-mini tester, the detection gap \emph{widens} to \spec{} $93\%$ vs \freeplus{} $53\%$
($+40$\pp{}), and grounding still cuts false alarms ($47\%\to35\%$), though a weak tester
keeps a precision floor (it mis-applies even a correct rule), consistent with the tester
sweep of \S\ref{app:e2}.

\begin{table}[tb]\centering
\caption{Standard-library-oracle slice: eight utilities reimplemented from under-specified
tickets under neutral names (15 buggy, 17 correct one-shots), judged against the real
CPython function. The ungrounded arms invent wrong expectations for genuine library
behaviour (high false alarms); a coverage-maximising prompt covers paths but misses the
contract.}
\label{tab:std}
\begin{tabular}{llccc}
\toprule
Tester & arm & detection & false alarm & tester acc\\
\midrule
Claude Sonnet 4.6 & \freeplus{}         & 83\% (25/30)  & 68\% (23/34) & 80\%\\
Claude Sonnet 4.6 & \spec{}             & 100\% (30/30) & 0\% (0/34)   & 94\%\\
Claude Sonnet 4.6 & coverage-maximising & 37\% (11/30)  & 0\% (0/34)   & 94\%\\
\midrule
GPT-4o-mini & \freeplus{}         & 53\% (16/30)  & 47\% (16/34) & 83\%\\
GPT-4o-mini & \spec{}             & 93\% (28/30)  & 35\% (12/34) & 86\%\\
\bottomrule
\end{tabular}
\end{table}

\begin{figure}[tb]\centering
\begin{tikzpicture}
\begin{axis}[width=0.92\textwidth, height=4.6cm, ybar, bar width=6pt,
  ymin=0, ymax=112, ytick={0,25,50,75,100}, ylabel={detection (\%)},
  symbolic x coords={core,b,rw,div,n,std,std-weak,repo},
  xtick=data, x tick label style={font=\footnotesize},
  ylabel style={font=\small}, legend pos=south west,
  legend style={font=\footnotesize, fill=white, fill opacity=0.92, draw=codeframe, draw opacity=1},
  area legend, enlarge x limits=0.08,
  major grid style={draw=plotgrid, line width=0.35pt}, ymajorgrids]
\addplot[fill=cfreeplus, draw=cfreeplus!60!black] coordinates {(core,60)(b,57)(rw,32)(div,83)(n,96)(std,83)(std-weak,53)(repo,40)};
\addplot[fill=cspec, draw=cspec!60!black] coordinates {(core,100)(b,95)(rw,100)(div,100)(n,100)(std,100)(std-weak,93)(repo,90)};
\legend{\freeplus{}, \spec{}}
\end{axis}
\end{tikzpicture}
\caption{Detection (buggy one-shots caught) by arm across the study's slices: the main
core set and the independent-stack slices (\texttt{b}, \texttt{rw}, the \texttt{div} and
\texttt{n} slices of App.~\ref{app:scale2}, the \texttt{std} slice and its weak-tester
variant, and \texttt{repo}, the real transformers/NLTK oracle of App.~\ref{app:repo}).
\spec{} (blue) reaches or nears $100\%$ on every slice; \freeplus{} (orange) varies with
how obvious each slice's edges are, but is never higher.}
\label{fig:allslices}
\end{figure}
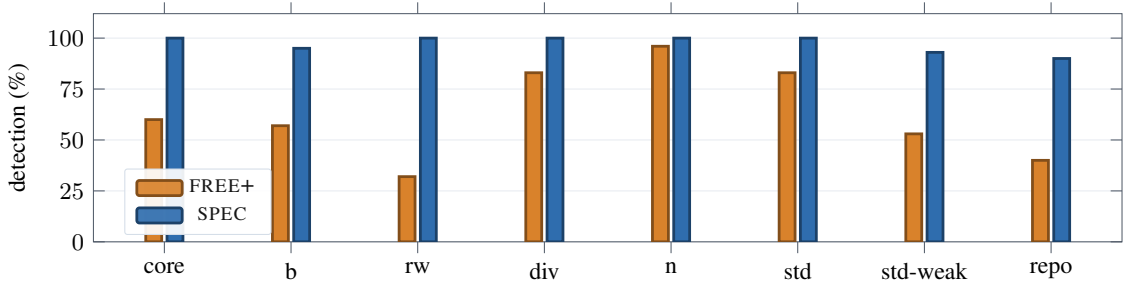

\paragraph{A pooled task-level test.} Across all $36$ independent-stack
specification-completeness tasks that produced a buggy one-shot (\S\ref{app:b},
\S\ref{app:rw}, \S\ref{app:scale2}, this slice, App.~\ref{app:repo}, and App.~\ref{app:pkg}), \spec{}
detection is $\ge$ \freeplus{} on \emph{every} task, strictly greater on $14$ and tied on
$22$; a one-sided sign test gives $p\approx6\times10^{-5}$. This complements the main
study's task-level test (final correctness, 18 tasks, $p=0.002$) with a detection-based
test on an independent stack, and Figure~\ref{fig:allslices} shows the detection gap
slice by slice.

\subsection{Real third-party repositories as the oracle: transformers and NLTK}\label{app:repo}
The standard-library slice uses CPython as the oracle; we push the same idea to real,
widely-used \emph{machine-learning} repositories. We take seven small pure-Python utilities
from HuggingFace \texttt{transformers} (the BERT tokenizer's \texttt{whitespace\_tokenize},
\texttt{\_run\_strip\_accents}, \texttt{\_run\_split\_on\_punc}, \texttt{\_clean\_text}, and
the \texttt{\_is\_punctuation} classifier) and NLTK (\texttt{edit\_distance};
\texttt{reduce\_lengthening}), vendor each \emph{verbatim} at a pinned commit (transformers
\texttt{c21da1b5}, NLTK \texttt{16a32d68}), and use that real function as the oracle. A coder
reimplements each from an under-specified, neutrally-named ticket; the one-shot bug rate is
$36\%$, concentrated on the genuinely edge-laden tokenizer functions
(\texttt{\_clean\_text}'s control-versus-whitespace rule, \texttt{\_is\_punctuation}'s
ASCII-range definition, punctuation splitting), while textbook utilities like
\texttt{edit\_distance} are reproduced correctly.

Against this real-repository oracle the grounding gap is large (Table~\ref{tab:repo}):
\spec{} detects $90\%$ of the bugs versus \freeplus{}'s $40\%$, a $+50$\pp{} gap, with no
false alarms on either arm. The ungrounded tester writes accurate but shallow cases that
exercise the happy path without probing the library's documented edge contracts, so it
misses the deviations that matter; the enumerated rules, restating the repository's own
behaviour, aim tests straight at them. These tasks are included in the pooled task-level
test above and in Figure~\ref{fig:allslices}.

\begin{table}[tb]\centering
\caption{Real third-party ML-repository oracle: seven pure-Python utilities vendored
verbatim from HuggingFace transformers and NLTK at pinned commits (28 one-shots, 10 buggy,
18 correct), judged against the real functions. \spec{} detects $+50$\pp{} more of the bugs
than \freeplus{}.}
\label{tab:repo}
\begin{tabular}{lccc}
\toprule
arm & detection & false alarm & tester acc\\
\midrule
\freeplus{} & 40\% (8/20)  & 0\% (0/36) & 100\%\\
\spec{}     & 90\% (18/20) & 0\% (0/36) & 95\%\\
\bottomrule
\end{tabular}
\end{table}

\subsection{A real package-index library as the oracle: PyPA \texttt{packaging}}\label{app:pkg}
The transformers/NLTK slice shows ungrounded testing \emph{missing} bugs. A real library can
also draw out the opposite failure: an ungrounded tester that guesses the under-specified
edges \emph{wrong} and so builds an oracle that rejects correct and buggy code alike. We take
four version-handling utilities from the PyPA \texttt{packaging} library (name
canonicalisation, version canonicalisation, version comparison, and version validity) and use
the \emph{real installed library} (version 26.2) as the oracle, with the gold inputs and
expected values copied verbatim from \texttt{packaging}'s own test suite at a pinned commit
(\texttt{e80f70f8}). A coder reimplements each from an under-specified, neutrally-named ticket.
The bug rate is high ($14/16$): the tickets name the common case but leave PEP~440's edges
(non-zero epochs, local versions, pre/post/dev tags, a leading \texttt{v}, and separator and
newline handling) implicit, so the coders write naive dotted-integer logic that misses them.

On this slice detection saturates (both arms catch all $14$ buggy one-shots), so the arms
separate on \emph{precision} instead (Table~\ref{tab:pkg}). Asked to test edges the ticket
never pins down, \freeplus{} writes expected values that disagree with the real library $15\%$
of the time; for instance it assumes an empty or malformed version string raises, when
\texttt{packaging} returns it unchanged. A suite with wrong expected values fails every
candidate it judges, so \freeplus{} flags both correct one-shots as buggy (a $100\%$ false
alarm rate) and cannot tell correct code from buggy at all. Restating the library's own rules
fixes every expected value ($100\%$ tester accuracy), so \spec{} keeps the full detection while
dropping false alarms to $0\%$: it accepts both correct candidates and rejects all fourteen
buggy ones. On a real third-party oracle, grounding turns a suite that rejects everything into
one that classifies perfectly. These four tasks are included in the pooled task-level test
above.

\begin{table}[tb]\centering
\caption{Real package-index-library oracle: four version utilities reimplemented against the
installed PyPA \texttt{packaging} library (version 26.2; gold data from the project's own tests
at commit \texttt{e80f70f8}). $16$ one-shots ($14$ buggy, $2$ correct). Detection is at ceiling
for both arms; the gap is in precision: \freeplus{} writes wrong expected values and rejects
every candidate, while \spec{} accepts the correct ones and rejects the buggy ones.}
\label{tab:pkg}
\begin{tabular}{lccc}
\toprule
arm & detection & false alarm & tester acc\\
\midrule
\freeplus{} & 100\% (28/28) & 100\% (4/4) & 85\%\\
\spec{}     & 100\% (28/28) & 0\% (0/4)   & 100\%\\
\bottomrule
\end{tabular}
\end{table}

\end{document}